\documentclass[12pt]{article}
\usepackage[margin=1in]{geometry}
\usepackage{setspace}
\usepackage{authblk}

\usepackage{amsmath,amssymb}
\usepackage{booktabs}
\usepackage{graphicx}
\usepackage{longtable}
\usepackage{array}
\usepackage{rotating}
\usepackage{tikz}
\usetikzlibrary{positioning,arrows.meta,calc}
\usepackage{enumitem}

\usepackage{xr-hyper}
\usepackage[hidelinks]{hyperref}

\usepackage{natbib}
\bibpunct[, ]{(}{)}{,}{a}{}{,}

\let\parencite\citep
\let\textcite\citet

\newcommand{\sym}[1]{\ifmmode^{#1}\else\(^{#1}\)\fi}

\newcommand{\feLPMcoeffVal}{-0.0875}
\newcommand{\creXTape}{ 0.0137}
\newcommand{\creXTapeSE}{ 0.0023}

\newcommand{\iptwCREape}{ 0.0705}
\newcommand{\iptwCREapeSE}{ 0.0032}

\newcommand{\apeClickPP}{ 7.1}
\newcommand{\apeReportPP}{ 0.8}
\newcommand{\apeSafePP}{ 0.8}

\newcommand{\jaccardIntCoeff}{ 0.178}

\newcommand{\desEmotCoeff}{-0.243}
\newcommand{\desPitchCoeff}{-0.170}
\newcommand{\desLandCoeff}{-0.122}
\newcommand{\desEmailCoeff}{ 0.090}
\newcommand{\desJointChi}{125.40}

\newcommand{\cueJointAuth}{-0.150}
\newcommand{\cueJointUrg}{ 0.275}

\newcommand{\cueJointCur}{-0.402}

\newcommand{\cueJointTransTemp}{ 0.426}

\newcommand{\cueJointChi}{162.54}

\newcommand{\engWaldChi}{ 7.93}
\newcommand{\engWaldP}{ 0.0049}

\newcommand{\engDisengApe}{ 0.0690}

\newcommand{\engEngApe}{ 0.0801}

\newcommand{\engDisengApePct}{ 6.9}
\newcommand{\engEngApePct}{ 8.0}
 
\DeclareRobustCommand{\yy}{\tikz[baseline=-0.5ex]{\fill[black!80] (0,0) circle (2.5pt);}}  \DeclareRobustCommand{\nn}{\tikz[baseline=-0.5ex]{\fill[gray!20] (0,0) circle (0.8pt);}} 

\newcommand{\tacticbar}[5]{\tikz[baseline=0.5ex, x=0.15cm, y=0.15cm]{
        \draw[gray!30, line width=0.5pt] (-0.5, 0) -- (4.5, 0);
        \foreach \val [count=\i] in {#1, #2, #3, #4, #5} {
            \ifnum\val=1
                \fill[black!85] (\i-1-0.2, 0) rectangle (\i-1+0.2, 2.2);
            \else
                \fill[black!15] (\i-1-0.2, 0) rectangle (\i-1+0.2, 0.4);
            \fi
        }
    }}

\begin{document}

\title{\textit{Research Note}---Breaking Bad Email Habits: Bounding the Impact of Simulated Phishing Campaigns}

\author[1]{Muhammad Zia Hydari}
\author[2]{Idris Adjerid}
\author[3]{Yingda Lu}
\author[1]{Narayan Ramasubbu}

\affil[1]{School of Business, University of Pittsburgh \\ \texttt{hydari@alum.mit.edu, narayanr@pitt.edu}}
\affil[2]{Pamplin College of Business, Virginia Polytechnic Institute and State University \\ \texttt{iadjerid@vt.edu}}
\affil[3]{University of Illinois at Chicago \\ \texttt{yingdalu@uic.edu}}

\date{}
\maketitle

\begin{abstract}
Simulated phishing campaigns are among the most widely deployed tools for reducing organizational cyber risk. Yet the behavioral data these campaigns produce have an underappreciated structural feature and a resulting complication: because training is triggered by clicking, the very employees who receive intervention are those who have already demonstrated susceptibility. This endogeneity, combined with the difficulty of separating genuine habit formation from stable individual differences in vulnerability, means that standard analyses of simulated phishing logs can substantially mischaracterize whether and for whom these programs work. In this \textit{Research Note}, we develop a generalizable analytic framework that addresses both biases simultaneously. We utilize marginal structural models (MSMs) to correct for the endogenous, click-triggered assignment of training, while integrating correlated random effects (CRE) for dynamic nonlinear panels to disentangle true state dependence from stable employee heterogeneity. The combined MSM+CRE estimator, demonstrated on de-identified logs from 17 campaigns delivered to university faculty and staff (192,840 person-by-campaign observations), reveals that analyses ignoring stable differences overstate the causal persistence of clicking; most observed repeat clicking reflects who employees are, not what recent failures have done to them. This persistence is strongly context dependent; it amplifies when successive campaigns share similar persuasion cues. Teachable-moment design features also matter: emotion or heuristic framing and explicit reporting pitch can largely eliminate persistence, while annotated-email cues modestly exacerbate it. Finally, employees who engage with the education page still exhibit greater persistence than those who dismiss it, consistent with an emboldening mechanism in consequence-free training environments. We contribute to the IS security literature, methodologically, by integrating MSMs and CRE into a portable analytical framework that any organization with standard phishing-simulation platform logs can apply; and practically, by identifying specific education-design levers and strategies for managing a phishing-test paradox in organizations, so that they can more effectively design, sequence, and evaluate their simulated phishing programs.
\end{abstract}


\vspace{1em}

\section{Introduction}

Phishing remains a persistent organizational threat because attackers can cheaply tailor deceptive messages that exploit employees' limited attention, time pressure, and routine email processing. Organizations, therefore, invest heavily in simulated phishing campaigns and associated security-awareness training. These programs create repeated, naturalistic opportunities for employees to encounter and handle suspicious messages. They also generate detailed behavioral logs that can, in principle, reveal whether employees learn from experience and feedback.

Despite their widespread adoption, analyzing and extracting reliable insights from simulated phishing data is challenging. These empirical challenges, often inadequately addressed in research and practice, distort conclusions about whether simulated phishing campaigns reduce risk. First, when the same employees are tracked across multiple campaigns, observed patterns can reflect both genuine learning (or lack thereof) and persistent susceptibility stemming from habitual routines. Methods developed in much of the empirical phishing literature remain cross-sectional, involve single exposure, or examine effects over short horizons, which are inadequate to shed light on longer-horizon, within-person dynamics in real-world settings \parencite{Moody2017WhichPhish, Goel2017GotPhished, Wright2023Context}. 

Second, the most common intervention mechanism in phishing-awareness training is click contingent. In this scheme, when an employee clicks on a link in a simulated phishing email, they are redirected to an educational ``teachable-moment'' page, turning each failure into an on-the-spot training opportunity. Consequently, clicking is both an outcome (susceptibility) and a trigger for treatment (training). Treatment assignment is, therefore, endogenous to past behavior, and standard panel models fail on two fronts: fixed-effects estimators suffer from severe Nickell bias in short dynamic panels \parencite{Nickell1981Bias}, while pooled estimators yield biased estimates because time-varying confounders are affected by prior treatment \parencite{Robins2000MSM, Cole2008IPTW}. That is, comparing employees who received training because they clicked with others who did not, confounds the effect of the training itself with the prior susceptibility that led to clicking in the first place. Together, these two challenges mean that a seemingly straightforward question, ``\textit{does clicking a simulated phish and receiving click-triggered feedback make an employee more or less likely to click on the next one?},'' cannot be answered reliably with conventional methods. Naive analyses risk conflating real behavioral change with statistical bias, potentially leading organizations to misjudge the effectiveness of their security-awareness programs.

In addition, there is an implicit intervention design challenge in repeated-exposure phishing panels. Observed click persistence is ambiguous. When the same employees click repeatedly across campaigns, this pattern conflates two distinct mechanisms. It could reflect \textit{true state dependence}, where the act of clicking and experiencing training causally alters the employee's future behavior (i.e., genuine habit formation or learning). Alternatively, it could merely reflect \textit{stable individual heterogeneity}, where certain employees possess enduring baseline traits—such as lower technical literacy or high-pressure roles—that make them consistently more click-prone regardless of past feedback. Distinguishing between the two is important for intervention design: state dependence implies downstream benefits from preventing a click episode, while stable heterogeneity implies the need for sustained targeting and complementary technical safeguards.

This \textit{Research Note} develops and applies a generalizable analytic framework that addresses these challenges and disentangles the confounding effects, enabling credible causal inference from the behavioral logs that simulated phishing campaigns collect. The framework examines how employees adapt after click-contingent feedback and when that feedback is most effective. To illustrate our methodology, we study de-identified event logs from a large research university that deployed the Cofense PhishMe platform to deliver simulated phishing emails to faculty and staff from June 2016 through February 2020. Using an employee-by-campaign email instance as the unit of analysis, we examine 192,840 observations pertaining to 19,341 employees across 17 simulated phishing campaigns. For each delivered phish message, the logs record clicking and reporting behaviors (and, for subsets of campaigns, \textit{opens} and deeper compromise actions).

\begin{figure}[htbp]
\centering
\begin{tikzpicture}[node distance=1.6cm,>=Latex]
  \node[draw,rounded corners,align=center,inner sep=6pt] (tminus) {Exposure $t-1$\\\footnotesize Click$_{t-1}$, Report$_{t-1}$};
  \node[draw,rounded corners,align=center,inner sep=6pt,right=2.4cm of tminus] (t) {Exposure $t$\\\footnotesize Click$_t$};
  \node[draw,rounded corners,align=center,inner sep=6pt,right=2.4cm of t] (tplus) {Next exposure $t+1$\\\footnotesize Click$_{t+1}$, Report$_{t+1}$};

  \node[draw,rounded corners,align=center,inner sep=6pt,below=1.1cm of t] (tm) {Teachable-moment\\page shown if Click$_t=1$};

  \draw[->] (tminus) -- node[above]{\footnotesize history} (t);
  \draw[->] (t) -- node[above]{\footnotesize causal effect} (tplus);
  \draw[->] (t) -- node[right]{\footnotesize treatment delivery} (tm);
  \draw[->,dashed] (tminus) .. controls +(2.2,0.7) and +(-2.2,0.7) .. node[above]{\footnotesize time-varying confounding} (tplus);
\end{tikzpicture}

\caption{Repeated exposures with click-contingent training. Clicking at exposure $t$ both indicates susceptibility and triggers the teachable-moment page. Histories affect both treatment and subsequent outcomes, motivating marginal structural models.}
\label{fig:timeline}
\end{figure}
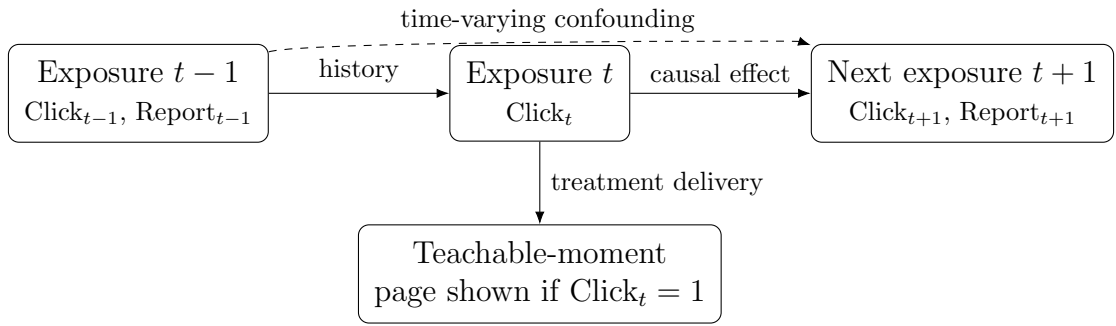

Our analysis setting is characterized by a structural feature that is ubiquitous in practice but rarely modeled explicitly. As shown in Figure~\ref{fig:timeline}, when an employee clicks in campaign $t$, the platform shows an educational page immediately after the click. We conceptualize this exposure as a time-varying treatment, denoted  \text{Click}$_t$, because it captures both the behavioral failure and the embedded, click-triggered feedback. The causal question is then forward looking: how does a click and its associated teachable moment in campaign $t$ affect behavior in the employee's next observed exposure $(t+1)$?

To explicitly resolve these empirical challenges, we develop an analysis framework that addresses both biases simultaneously. First, to correct for the endogenous, click-triggered assignment of training, we apply marginal structural models (MSMs) estimated via stabilized inverse probability of treatment weights (IPTW) \parencite{Robins2000MSM, Cole2008IPTW}. This corrects for the time-varying confounding introduced when an employee's prior behaviors dictate their subsequent treatment exposure. Second, to resolve the ambiguity between true state dependence and stable employee heterogeneity within a dynamic binary panel, we adopt a correlated random effects (CRE) approach for nonlinear panels with initial conditions \parencite{Wooldridge2005InitialConditions}. Finally, our primary estimator unites these solutions into a single, combined MSM+CRE specification: a pooled probit containing CRE terms that is estimated using the stabilized IPTW, with average partial effects (APEs) reported for straightforward interpretability.

We complement the main estimates with three extensions. First, we test contextual transfer by measuring cue similarity between scenarios at $t$ and $t+1$ using a Jaccard index and estimating whether similarity amplifies persistence. Second, we examine teachable-moment page design as a moderator of persistence, focusing on design features observed on the post-click education page (annotated cues, reporting pitch, and emotion or heuristic framing). Third, we decompose persistence by phishing cue type and by engagement with the education page to test whether persistence operates through cue-specific vulnerability channels and whether consequence-free engagement is consistent with an emboldening mechanism.

The results of the illustrative analysis using our approach on the campaign data support four conclusions. First, even after jointly adjusting for the dynamic feedback of prior training (time-varying confounding) and employees' enduring baseline risks (stable heterogeneity), clicking at $t$ increases the probability of clicking at $t+1$ by about \apeClickPP\ percentage points, while increasing reporting and safe handling by about \apeReportPP\ and \apeSafePP\ points.\footnote{Online Appendix \S \ref{EC-app:sec:ec_safe_paradox} provides the mathematical intuition for this result, illustrating how safe handling and clicking can increase simultaneously.} Second, an analytical progression reveals that most naively estimated persistence reflects stable heterogeneity rather than true state dependence, and the state-dependence component is bounded between about 1.4 and \apeClickPP\ percentage points. Third, persistence is strongly context dependent: cue similarity and cue type jointly shape whether clicking at $t$ is emboldening or self-correcting. Fourth, education design can eliminate persistence when it increases psychological salience (emotion or heuristic framing) and makes defensive action scripts concrete (reporting pitch). Conversely, consequence-free engagement with weaker designs can inadvertently reinforce risky behavior, creating a \textit{phishing-test paradox} where the intervention emboldens the susceptibility it aims to cure.

This \textit{Research Note} contributes to the Information Systems security literature in three key ways. First, it advances a robust analytical framework for examining data from repeated exposures, long-horizon phishing simulation campaigns. For this methodological contribution, we integrated MSMs that address time-varying confounding using stabilized IPTW \parencite{Robins2000MSM, Cole2008IPTW} with correlated random effects for dynamic nonlinear panels, which address stable heterogeneity \parencite{Wooldridge2005InitialConditions}. This approach yields more credible estimates of true state dependence and a transparent analytical progression across estimators. Substantively, our illustrative analysis using a real-world dataset provides within-person causal evidence on how susceptibility and response evolve under repeated simulated attacks, complementing cross-sectional correlates of susceptibility \parencite{Vishwanath2011WhyDoPeople, Moody2017WhichPhish, Goel2017GotPhished}. Conceptually, the application of our framework highlights contextual transfer and teachable-moment design as determinants of whether learning generalizes across repeated phishing campaigns. 

\section{Literature Review and Research Gaps}
\label{sec:litreview}

\subsection{Phishing susceptibility, information processing, and context}

Research on phishing has documented that employees and consumers rely on heuristics and cognitive shortcuts when processing email and online requests. Integrated information-processing models emphasize that susceptibility reflects the interaction of individual differences, attention allocation, and message cues \parencite{Vishwanath2011WhyDoPeople, Wang2016Overconfidence}. Field studies and large-scale logs confirm that clicking risk is concentrated in a subset of users, varies by organizational role and context, and can be shaped by workplace constraints \parencite{Goel2017GotPhished, WILLIAMS20181}. More recently, multilevel evidence suggests that susceptibility depends on situational context and organizational setting, not only on stable individual traits \parencite{Wright2023Context}. Crucially, employees rarely process emails in isolation. Real-world multitasking and high working memory load severely impair detection performance, as the cognitive demands of the workplace often override active security goals \parencite{Lu02012026}.

This literature provides important correlational insights, but it leaves open a central dynamic question that is consequential for both theory and practice: how does susceptibility change \emph{within the same individuals} across repeated exposures to heterogeneous phishing scenarios? When longitudinal data are available, analyses often emphasize prediction (who will click next) rather than causal estimation of feedback or training effects \parencite{Abbasi2021PhishingFunnel}. Yet in operational settings, employees are repeatedly exposed to campaigns, and their histories of clicks, reports, and feedback become part of the causal process, rendering current approaches insufficient for reliably evaluating whether training works.

\subsection{Interventions: embedded training, teachable moments, and reporting}

A second stream studies interventions that seek to reduce susceptibility and improve protective behaviors. Embedded training approaches provide instruction at the moment of failure, on the premise that feedback is more salient when it follows an error \parencite{Kumaraguru2007PhishingTraining}. Recent work has evaluated interventions such as mindfulness-based training \parencite{Jensen2017Mindfulness}, susceptibility claims \parencite{Jensen2021Claims}, and gamified reporting incentives \parencite{Jensen2022Gamification}. These interventions can improve knowledge and behavior, but effects may attenuate over time and may fail to transfer across changing cues and contexts. Moreover, reporting has emerged as a critical complement to click avoidance because it enables rapid incident response and collective defense. Reporting systems and warning mechanisms can affect both detection and user decision making \parencite{Nguyen2021Crowdsourced}. Another stream of related studies clarifies how experiential exposure and feedback shape phishing-related learning. For example, \textcite{Wright2010Experiential} emphasize experiential learning mechanisms in anti-phishing education, suggesting that the form and timing of feedback can influence whether users update their decision rules broadly or instead encode scenario-specific lessons. Complementary work highlights the value of technical defenses that can augment user-facing training. Furthermore, \textcite{Abbasi2015EnhancingPredictive} show that exploiting website similarity can improve predictive analytics for anti-phishing, reinforcing the view that effective defense typically requires both human and technical layers.

A central methodological challenge that remains is that in many organizations, training exposure is endogenous. Employees often receive training precisely because they engaged in risky behavior. Simple comparisons of trained and untrained users therefore confound the effect of training with underlying risk propensity and with the influence of prior outcomes. Such bias can lead organizations to misjudge the effectiveness of their training programs and to misallocate resources in response. 

\subsection{Message cues and contextual transfer}

A third stream highlights that phishing success depends on social-engineering tactics and the manipulation of message cues. Influence techniques such as authority, urgency, and financial appeals can increase compliance, yet their effects vary across populations and work settings \parencite{Wright2014InfluenceTechniques, WILLIAMS20181}. This variation suggests that learning and resistance may be cue contingent. If employees learn primarily to recognize specific cues or cue combinations, then transfer to later campaigns should depend on whether those scenarios share the same cues.

The notion of contextual transfer is consistent with evidence from sequential attacks in adjacent domains. For example, social-network phishing studies find that prior compliance predicts later compliance in multi-stage attacks, indicating that vulnerability can persist and compound across exposures \parencite{Vishwanath2017SocialMedia}. However, most phishing field studies do not measure whether similarity across stimuli strengthens or weakens this persistence.

\subsection{Research gaps}

Three gaps emerge from our synthesis. First, we lack strong within-person causal evidence on how click-triggered feedback affects subsequent behavior in repeated-exposure panels when treatment assignment depends on prior outcomes and histories. Second, we know little about contextual transfer across heterogeneous campaigns, including whether similarity in cues amplifies persistence among prior clickers. Third, the teachable-moment pages shown after clicks vary meaningfully in design, yet existing studies rarely exploit this heterogeneity to identify which design elements attenuate susceptibility and which strengthen reporting.

To facilitate a transparent synthesis of these gaps, the Online Appendix reports structured summaries of prior phishing studies across susceptibility, message design, interventions, and detection, including papers cited in this section (online appendix \S \ref{EC-app:sec:ec_lit}). The Online Appendix also reports additional scenario-coding detail and full scenario similarity matrices (online appendix \S \ref{EC-app:sec:ec_coding} and \S \ref{EC-app:sec:ec_similarity}). While the appendix provides this extended context, the remainder of this \textit{Research Note} focuses on the analytical framework and causal design required to address the dynamic, endogenous nature of click-contingent feedback.

\section{Research Context, Data, and Measures}
\label{sec:data}

\subsection{Setting and panel construction}

We analyze de-identified logs exported from a simulated phishing platform deployed at a large research university. The platform delivered repeated simulated phishing emails to faculty and staff and recorded behaviors such as clicking and reporting. Each campaign corresponds to a specific scenario consisting of a phishing email template and, conditional on clicking, a teachable-moment educational page. We use \emph{campaign} to denote the time-ordered deployment and \emph{scenario} to denote the underlying content template; in our data there is a one-to-one mapping between campaigns and scenarios, but we reserve ``scenario'' for content coding and ``campaign'' for the temporal index.

The cleaned analytic dataset contains 192,840 employee-by-campaign observations for 19,341 unique employees spanning 17 campaigns from June 7, 2016 through February 11, 2020. Employees appear in multiple campaigns, with a mean of 9.97 campaigns per employee and a median of 11 (Table~\ref{tab:descriptives}). For each employee $i$, we order exposures chronologically and define $t$ as the within-person exposure index. Our estimands focus on transitions from exposure $t$ to the employee's next observed exposure $(t+1)$.  Accordingly, the MSM transition models are estimated on the subset of exposure instances with an observed next exposure (173{,}499 transitions), whereas descriptive statistics use the full 192{,}840 employee-by-campaign exposure instances.

\begin{table}[htbp]
\centering
\small
\caption{Sample description and key behavioral rates}
\label{tab:descriptives}
\begin{singlespace}
\begin{tabular}{l r}
\toprule
\textbf{Panel A: Sample} & \\
\midrule
User-by-campaign email instances & 192,840 \\
Unique employees (de-identified) & 19,341 \\
Distinct simulated phishing campaigns (scenarios) & 17 \\
Observation window (first to last campaign) & June 7, 2016 to February 11, 2020 \\
Mean campaigns per employee & 9.97 \\
Median campaigns per employee (25th, 75th) & 11 (5, 15) \\
\addlinespace
\textbf{Panel B: Baseline employee composition} & \\
\midrule
Faculty & 20.4\% \\
Staff & 46.5\% \\
Role not observed in administrative fields & 33.1\% \\
\addlinespace
\textbf{Panel C: Behavioral outcomes} & \\
\midrule
Click rate (all campaigns) & 12.47\% \\
Report rate (all campaigns) & 5.62\% \\
Open rate (when open tracking enabled) & 35.15\% \\
Credential submission rate (when applicable) & 2.99\% \\
\addlinespace
\textbf{Panel D: Raw transition probabilities (unweighted)} & \\
\midrule
$\Pr(\text{Click}_{t+1}=1 \mid \text{Click}_{t}=1)$ & 20.04\% \\
$\Pr(\text{Click}_{t+1}=1 \mid \text{Click}_{t}=0)$ & 9.25\% \\
$\Pr(\text{Report}_{t+1}=1 \mid \text{Report}_{t}=1)$ & 42.63\% \\
$\Pr(\text{Report}_{t+1}=1 \mid \text{Report}_{t}=0)$ & 4.11\% \\
\bottomrule
\addlinespace[0.2cm]
\end{tabular}
\end{singlespace}
\begin{flushleft}\footnotesize
\textit{Notes:} The unit of observation is one simulated phishing email delivered to one employee in one campaign. Open and credential-submission measures are scenario dependent; reported rates are computed over nonmissing observations. Transition probabilities are computed over adjacent exposures for the same employee and are intended for descriptive motivation; causal estimates are reported in subsequent tables.
\end{flushleft}
\end{table}
 
\subsection{Outcomes}

\paragraph{Clicking.}  \text{Click}$_t$ equals 1 if the employee clicked the simulated link in campaign $t$. In our setting,  \text{Click}$_t=1$ also implies immediate exposure to the post-click teachable-moment page.

\paragraph{Reporting.}  \text{Report}$_t$ equals 1 if the employee reported the message as suspicious using the platform's reporting mechanism.

\paragraph{Safe handling.} We define  \text{Safe}$_{t}=1$ if the employee reports at $t$ without clicking at $t$, that is,  \text{Report}$_{t}=1$ and  \text{Click}$_{t}=0$. This captures a conservative safe-response behavior that is operationally meaningful for incident response.

Clicking and reporting are not mutually exclusive in this setting: an employee may click and later report the same message. Accordingly, we model  \text{Click}$_{t}$ and  \text{Report}$_{t}$ as separate binary outcomes at $t$, and the safe-handling outcome isolates the joint event of reporting without clicking at $t$.

\subsection{Covariates and histories}

Because click-triggered feedback is endogenous to prior behavior, we construct pre-treatment histories for each exposure. These include cumulative prior clicks and reports, lagged click and report indicators, the number of prior exposures, and the time gap since the preceding exposure. We also include baseline covariates measured at the employee's first observed exposure, including role indicators (faculty, staff, role unknown), job-status categories, organizational unit categories, and tenure since account creation with a missingness indicator.

\subsection{Scenario-level coding scheme and similarity measures}
\label{sec:scenario_coding}

To characterize each campaign and its teachable-moment page, we coded scenario-level content features from screenshots of the phishing email and post-click educational page. The coding scheme captures two layers: (i) the persuasion and context cues embedded in the phishing email (the stimuli that shape information scent) and (ii) the pedagogical and design features of the teachable-moment page (the intervention content shown only to clickers).  Scenario coding was performed using a pre-specified codebook and screenshots of each phishing email and its post-click teachable-moment page. One author conducted the initial coding for all 17 scenarios, and the remaining authors reviewed the codes against the underlying screenshots and discussed any ambiguities until reaching a consensus set of indicators used in the empirical models. The Online Appendix \S \ref{EC-app:sec:ec_coding} reports the full codebook and scenario-level codes.

For phishing emails, we code five social-engineering tactic cues: authority, urgency, financial appeal, curiosity, and internal context. We also code two email-format indicators that capture whether the phishing email uses a transactional template (button-style calls to action) and whether it uses an attachment lure.  For teachable-moment pages, we code whether the page includes annotated email cues, annotated landing cues (including credential-entry cues), an explicit reporting pitch, and emotion or heuristic framing.  Table~\ref{tab:scenario_coding} summarizes the coded features across the 17 scenarios.

\begin{table}[htbp]
\centering
\scriptsize
\caption{Scenario-level coding of phishing email and teachable-moment page characteristics.}
\label{tab:scenario_coding}
\renewcommand{\arraystretch}{1.5}
\resizebox{\textwidth}{!}{\begin{tabular}{rc ccccc c cc c ccccc}
\toprule
& & \multicolumn{5}{c}{Phishing-email tactic cues} && \multicolumn{2}{c}{Email format} && \multicolumn{5}{c}{Teachable-moment page features} \\
\cmidrule{3-7} \cmidrule{9-10} \cmidrule{12-16}
Scen. & Profile & Authority & Urgency & Financial & Curiosity & Internal &&
\begin{tabular}{@{}c@{}}Trans.\\Temp.\end{tabular} &
\begin{tabular}{@{}c@{}}Attach.\\Lure\end{tabular} &&
\begin{tabular}{@{}c@{}}Ann.\\Email\end{tabular} &
\begin{tabular}{@{}c@{}}Ann. LP/\\Cred.\end{tabular} &
\begin{tabular}{@{}c@{}}Report\\Pitch\end{tabular} &
\begin{tabular}{@{}c@{}}Emot./\\Heur.\end{tabular} &
\begin{tabular}{@{}c@{}}Scen.\\Theme\end{tabular} \\
\midrule
1 & \tacticbar{0}{0}{1}{1}{0} & \nn & \nn & \yy & \yy & \nn && \yy & \nn && \nn & \nn & \nn & \nn & \nn \\
2 & \tacticbar{0}{0}{1}{1}{0} & \nn & \nn & \yy & \yy & \nn && \yy & \nn && \nn & \nn & \nn & \nn & \nn \\
3 & \tacticbar{1}{1}{0}{0}{1} & \yy & \yy & \nn & \nn & \yy && \nn & \nn && \yy & \nn & \nn & \nn & \nn \\
4 & \tacticbar{1}{1}{0}{0}{1} & \yy & \yy & \nn & \nn & \yy && \nn & \nn && \yy & \nn & \nn & \nn & \nn \\
5 & \tacticbar{1}{1}{0}{0}{1} & \yy & \yy & \nn & \nn & \yy && \nn & \nn && \nn & \yy & \nn & \nn & \nn \\
\addlinespace
6 & \tacticbar{0}{0}{1}{1}{0} & \nn & \nn & \yy & \yy & \nn && \yy & \nn && \yy & \nn & \nn & \nn & \nn \\
7 & \tacticbar{1}{1}{1}{0}{0} & \yy & \yy & \yy & \nn & \nn && \nn & \nn && \yy & \nn & \nn & \nn & \yy \\
8 & \tacticbar{0}{1}{0}{1}{0} & \nn & \yy & \nn & \yy & \nn && \nn & \nn && \yy & \nn & \nn & \nn & \nn \\
9 & \tacticbar{1}{1}{0}{1}{0} & \yy & \yy & \nn & \yy & \nn && \nn & \nn && \yy & \nn & \yy & \nn & \nn \\
10 & \tacticbar{1}{1}{0}{1}{1} & \yy & \yy & \nn & \yy & \yy && \nn & \nn && \yy & \yy & \nn & \yy & \nn \\
\addlinespace
11 & \tacticbar{1}{0}{1}{0}{1} & \yy & \nn & \yy & \nn & \yy && \nn & \nn && \yy & \nn & \yy & \nn & \nn \\
12 & \tacticbar{0}{0}{1}{1}{0} & \nn & \nn & \yy & \yy & \nn && \yy & \nn && \yy & \nn & \yy & \nn & \nn \\
13 & \tacticbar{0}{0}{0}{1}{1} & \nn & \nn & \nn & \yy & \yy && \nn & \yy && \yy & \nn & \yy & \yy & \nn \\
14 & \tacticbar{1}{0}{0}{1}{1} & \yy & \nn & \nn & \yy & \yy && \nn & \nn && \yy & \yy & \nn & \yy & \nn \\
15 & \tacticbar{1}{1}{0}{1}{1} & \yy & \yy & \nn & \yy & \yy && \nn & \nn && \yy & \nn & \yy & \yy & \nn \\
\addlinespace
16 & \tacticbar{0}{0}{1}{1}{0} & \nn & \nn & \yy & \yy & \nn && \yy & \nn && \yy & \nn & \yy & \yy & \nn \\
17 & \tacticbar{0}{1}{0}{1}{1} & \nn & \yy & \nn & \yy & \yy && \nn & \nn && \yy & \nn & \yy & \yy & \nn \\
\bottomrule
\end{tabular}}
\end{table}
 
\paragraph{Cue similarity.} For any two scenarios $s$ and $s'$, we define the Jaccard similarity over phishing-email tactic cues as
\begin{equation}
 \text{Jaccard}(s,s') = \frac{|\mathbf{c}_s \cap \mathbf{c}_{s'}|}{|\mathbf{c}_s \cup \mathbf{c}_{s'}|},
\label{eq:jaccard}
\end{equation}
where $\mathbf{c}_s$ is the set of tactic cues present in scenario $s$. The Jaccard index ranges from 0 (no shared present cues) to 1 (identical cue sets). We use it because it emphasizes overlap in present cues, which is consistent with a cue-activation mechanism in which shared cues trigger similar decision heuristics. Section \ref{EC-app:sec:ec_similarity} in the Online Appendix reports the full similarity matrix and additional descriptive comparisons.

\section{Empirical Strategy and Identification}
\label{sec:msm}

\subsection{A naive fixed-effects benchmark and why it fails}

A natural benchmark is a within-employee specification that partials out stable employee traits. For outcome $Y_{i,t+1}$, consider the fixed-effects model
\begin{equation}
Y_{i,t+1} = \alpha_i + \psi\,\text{Click}_{i,t} + \gamma_{c(t+1)} + \varepsilon_{i,t+1},
\label{eq:fe_lpm}
\end{equation}
where $\alpha_i$ is an employee fixed effect, $\gamma_{c(t+1)}$ is a fixed effect for the next campaign scenario, and $\text{Click}_{i,t}$ is the employee's click indicator at the prior exposure. In the click equation, $\text{Click}_{i,t}$ is a lagged dependent variable because $Y_{i,t+1}=\text{Click}_{i,t+1}$. In our data, this fixed-effects linear probability model yields a negative persistence coefficient (Table~\ref{tab:progression}, Column~1), which is not credible in this setting.  This failure is expected. In dynamic panels with fixed effects and short time series, the within transformation induces correlation between the demeaned lagged outcome and the demeaned error, generating Nickell bias of order $1/T$ \parencite{Nickell1981Bias}. With a mean of roughly ten exposures per employee and binary outcomes, fixed-effects estimators are therefore unreliable for estimating click persistence. This motivates an approach that addresses stable heterogeneity without relying on fixed effects and that simultaneously addresses time-varying confounding created by click-triggered feedback.

\subsection{Two identification challenges}

Our setting presents two distinct identification challenges. First, \emph{stable individual heterogeneity}. Employees differ in baseline susceptibility and vigilance due to enduring traits and job environments. If these stable differences are correlated with clicking histories, persistence estimates can conflate true state dependence (a click causally increases future clicking) with stable heterogeneity (some employees are persistently click prone).  Second, \emph{time-varying confounding under feedback}. Clicking at $t$ triggers the teachable-moment page, so clicking is simultaneously an outcome and a treatment assignment rule. Histories such as cumulative prior clicks and reports, exposure order, and the time gap since the prior exposure affect both the probability of clicking at $t$ and outcomes at $t+1$. These histories are also affected by prior clicks and prior feedback, so conventional regression adjustment can be biased because time-varying confounders are themselves post-treatment variables \parencite{Robins2000MSM, Hernan2020CausalInference}.

We address stable heterogeneity using correlated random effects (CRE) for dynamic nonlinear panels and address time-varying (TV) confounding using marginal structural models (MSM) with stabilized inverse probability of treatment weights. We then combine the two in a single MSM+CRE estimator.

\subsection{Marginal structural models and stabilized IPTW}

Let $A_{it}=\text{Click}_{it}$ denote the time-varying treatment and let $Y_{i,t+1}$ denote an outcome at the next exposure ($\text{Click}_{i,t+1}$, $\text{Report}_{i,t+1}$, or $\text{Safe}_{i,t+1}$). Let $\bar{A}_{i,t-1}$ denote past treatment history and let $\bar{L}_{it}$ denote observed time-varying histories up to exposure $t$. The stabilized weight for exposure $t$ is
\begin{equation}
SW_{it} = \prod_{k=0}^{t} \frac{\Pr(A_{ik} \mid \bar{A}_{i,k-1}, \mathbf{X}_i)}{\Pr(A_{ik} \mid \bar{A}_{i,k-1}, \bar{L}_{ik}, \mathbf{X}_i)},
\label{eq:sw}
\end{equation}
where $\mathbf{X}_i$ denotes baseline covariates. In our data, $\mathbf{X}_i$ includes role indicators (faculty, staff, role unknown), job-status categories, organizational-unit groupings, and tenure at first observed exposure. The history vector $\bar{L}_{ik}$ includes lagged click and report indicators, cumulative prior clicks and reports, exposure order, and the gap in days since the prior observed exposure.

In our implementation, the numerator model includes baseline covariates and campaign indicators and omits the time-varying history terms; the denominator model includes the same baseline covariates and campaign indicators and adds the full set of time-varying histories. We estimate both models with pooled logistic regressions and compute $SW_{it}$ for each observation. Following best practice, we trim extreme weights at the 1st and 99th percentiles to reduce variance inflation. In our data, 1{,}928 observations are capped at the 1st-percentile cutoff and 1{,}929 at the 99th-percentile cutoff (3{,}857 of 192{,}840; about 2.0\%), using cutoffs 0.164 and 1.775. Weight summaries and a histogram are reported in the Online Appendix \S \ref{EC-app:sec:weights}.

Under sequential ignorability and positivity, MSMs estimate the causal effect of clicking and its click-triggered feedback by fitting weighted outcome regressions in the IPTW pseudo-population \parencite{Robins2000MSM, Cole2008IPTW}. In Table~\ref{tab:progression}, Column~(4) reports an IPTW-weighted probit. To maintain consistency with the Wooldridge CRE framework introduced below, we estimate the structural outcome model using a probit link function rather than the conventional logit link. This choice is purely structural and does not alter our substantive conclusions; standard logit MSM estimates are reported in the Online Appendix \S \ref{EC-app:sec:ec_msm_compare} and are consistent with our main results.

\subsection{Correlated random effects for stable heterogeneity}

Dynamic binary panels raise an initial conditions problem: unobserved employee-specific heterogeneity can be correlated with the lagged outcome (here, prior clicking), and naive random-effects assumptions can confound true state dependence with stable differences. We use the correlated random effects approach for dynamic nonlinear panels \parencite{Wooldridge2005InitialConditions}, which projects the unobserved effect onto observable functions of the covariates (a Mundlak or Chamberlain device \parencite{Mundlak1978Pooling}). Concretely, we include: (i) an initial-condition indicator, $\text{Click}_{i1}$ (click status at the employee's first observed exposure), (ii) individual means of time-varying covariates used in the persistence models (Mundlak terms), (iii) the number of observed exposures, and (iv) department-group and job-status controls. The Mundlak terms are constructed from covariates that vary across campaigns in our data, including scenario similarity measures (Jaccard) and the coded teachable-moment design indicators (annotated email, annotated landing, reporting pitch, emotion or heuristic framing). These terms relax the strict random-effects independence assumption and reduce bias from stable employee heterogeneity.

\subsection{The combined MSM+CRE estimator and bounding logic}
\label{subsec:combinedMSM.CRE}

Our primary estimator combines both adjustments in a pooled probit:
\begin{equation}
\Phi^{-1}\Pr(Y_{i,t+1}=1)=\alpha + \psi\,\text{Click}_{i,t} + \gamma_{c(t+1)} + \mathbf{h}_i'\boldsymbol{\delta},
\label{eq:iptwcre_main}
\end{equation}
where $\Phi(\cdot)$ is the standard normal CDF, $\gamma_{c(t+1)}$ are next-campaign fixed effects, and $\mathbf{h}_i$ collects the CRE terms described above. The dependent variable $Y_{i,t+1}$ is one of $\text{Click}_{i,t+1}$, $\text{Report}_{i,t+1}$, or $\text{Safe}_{i,t+1}$. The key regressor $\text{Click}_{i,t}$ equals 1 if employee $i$ clicked in the prior campaign and was therefore redirected to the teachable-moment page. We weight each observation by the trimmed stabilized IPTW and cluster standard errors by employee. We report average partial effects (APEs) to express persistence in percentage points.

This pooled probit with CRE terms is the practical vehicle for combining CRE and IPTW because standard random-effects probit routines (for example, \texttt{xtprobit, re} in Stata) do not accommodate the stabilized probability weights used in MSM estimation. Wooldridge's formulation shows that pooled probit with appropriate CRE terms provides consistent estimates of APEs for dynamic nonlinear panels under large-$N$, fixed-$T$ asymptotics \parencite{Wooldridge2005InitialConditions}. Because any finite CRE projection is necessarily approximate, we interpret MSM+CRE as our primary point estimate and as a plausible upper bound when residual stable heterogeneity remains. The corresponding CRE random-effects estimate provides a conservative lower bound. Accordingly, the true causal persistence plausibly lies between these two estimates.

\subsection{Analytical progression}

To clarify which identification threat each estimator addresses, we report an analytical progression table (Table~\ref{tab:progression}) that compares (i) a fixed-effects linear probability model (illustrating Nickell bias), (ii) an unweighted pooled probit, (iii) a CRE probit, (iv) an MSM probit (the MSM), and (v) MSM+CRE. This progression makes transparent how much of persistence is attributable to stable heterogeneity versus time-varying confounding.

\subsection{Extension specifications}

\paragraph{Cue similarity extension.} To test contextual transfer, we interact $\text{Click}_{i,t}$ with the Jaccard similarity between the phishing-email cue sets of campaigns $t$ and $t+1$ within the MSM+CRE probit:

\begin{equation}
\begin{aligned}
\Phi^{-1}\Pr( \text{Click}_{i,t+1}=1) &= \alpha + \psi\,\text{Click}_{i,t} \\
&\quad + \eta\,\text{Sim}_{i,t,t+1} \\
&\quad + \kappa (\text{Click}_{i,t} \times \text{Sim}_{i,t,t+1}) \\
&\quad + \gamma_{c(t+1)} + \mathbf{h}_i'\boldsymbol{\delta}.
\end{aligned}
\label{eq:iptwcre_sim}
\end{equation}

Here $\text{Sim}_{i,t,t+1}$ is the Jaccard similarity between the coded cue sets of the two scenarios, computed from the binary presence indicators for authority, urgency, financial appeal, curiosity, and internal context. A positive $\kappa$ indicates that persistence is amplified when the next scenario shares more cues with the prior one.

\paragraph{Teachable-moment design moderators.} We interact $\text{Click}_{i,t}$ with scenario-level indicators for teachable-moment design features in campaign $t$ (annotated landing, reporting pitch, emotion or heuristic framing, and annotated email). These features are observed in the screenshots and are coded at the scenario level (Table~\ref{tab:scenario_coding}). Because only clickers view the teachable-moment page, moderator main effects are not interpreted causally; interaction terms capture how the design of the click-triggered feedback changes the effect of a prior click on subsequent behavior.

\section{Results}
\label{sec:results}

\subsection{Analytical progression and main MSM+CRE estimates}

Table~\ref{tab:progression} reports an analytical progression that clarifies how alternative estimators address distinct identification threats. Column~(1) shows that a fixed-effects linear probability model yields a negative persistence estimate (\feLPMcoeffVal). Taken at face value, this would lead to erroneous conclusions about the efficacy of the training. However, this negative estimate is a known statistical artifact (Nickell bias) driven by the within-transformation in short dynamic panels, underscoring the danger of relying on standard fixed effects to evaluate simulated phishing campaigns. Column~(2) reports a pooled probit that addresses neither time-varying confounding nor stable heterogeneity and therefore overstates persistence. Columns~(3) and (4) show that addressing either stable heterogeneity (CRE probit) or time-varying confounding (MSM probit) reduces the implied persistence relative to Column~(2). Column~(5) combines both adjustments and is our primary specification.  

\begin{table}[htbp]
\centering
\caption{Analytical progression: identifying click persistence}
\label{tab:progression}
\begin{singlespace}
\def\sym#1{\ifmmode^{#1}\else\(^{#1}\)\fi}
\begin{tabular}{lccccc}
\toprule
&\multicolumn{1}{c}{(1)}&\multicolumn{1}{c}{(2)}&\multicolumn{1}{c}{(3)}&\multicolumn{1}{c}{(4)}&\multicolumn{1}{c}{(5)}\\
&\multicolumn{1}{c}{FE-LPM}&\multicolumn{1}{c}{Probit}&\multicolumn{1}{c}{CRE probit}&\multicolumn{1}{c}{MSM probit}&\multicolumn{1}{c}{MSM+CRE}\\
\midrule
Click\ensuremath{\sb{t}}&      -0.088\sym{***}&      &        &    &        \\
      &     (0.003)      &   &   &   &   \\\addlinespace
Click\ensuremath{\sb{t}}&   &       0.550\sym{***}&       0.093\sym{***}&       0.475\sym{***}&       0.367\sym{***}\\
&   &     (0.014)   &     (0.015)   &     (0.014)   &     (0.015)   \\\addlinespace
\midrule
\textbf{APE}   &    --   &      \textbf{0.1158}   &      \textbf{0.0137}   &      \textbf{0.0975}   &      \textbf{0.0705}   \\
Addresses   & --  &     Neither   & Stable het.   &TV confounding   &        Both   \\
Observations&     173,499   &     173,499   &     173,499   &     173,499   &     173,499   \\
\bottomrule
\addlinespace[0.2cm]
\end{tabular}
\end{singlespace}
\begin{flushleft}\footnotesize    \textit{Notes:} All columns report the coefficient on    \ensuremath{\text{Click}\sb{t}} from the indicated estimator.    Column~(1) is a fixed-effects linear probability model (subject to    Nickell bias). Columns~(2)--(5) are probit models; APEs are reported    at the foot of the table. CRE terms include initial condition    (\ensuremath{y\sb{i1}}), individual means of time-varying covariates    (Mundlak terms), exposure count, department group, and job status.    IPTW uses stabilized inverse probability of treatment weights trimmed at    the 1st and 99th percentiles. All models include next-campaign fixed    effects. Standard errors are clustered by employee.    \sym{*} \ensuremath{p<0.10}; \sym{**} \ensuremath{p<0.05};    \sym{***} \ensuremath{p<0.01}.    \end{flushleft}
\end{table}
 
Table~\ref{tab:iptwcre_main} reports the MSM+CRE main effects for clicking, reporting, and safe handling. The click-persistence effect remains positive and precisely estimated: clicking at exposure $t$ increases the probability of clicking at the next exposure by \iptwCREape\ (about \apeClickPP\ percentage points; s.e. \iptwCREapeSE). Reporting and safe handling increase by about \apeReportPP\ and \apeSafePP\ percentage points, respectively.\footnote{Online Appendix \S \ref{EC-app:sec:ec_safe_paradox} provides the mathematical intuition for this result, illustrating how safe handling and clicking can increase simultaneously.} These effects are economically modest but nontrivial given the low baseline rates of reporting and safe handling.

\begin{table}[htbp]
\centering
\caption{MSM+CRE main effects}
\label{tab:iptwcre_main}
\begin{singlespace}
\def\sym#1{\ifmmode^{#1}\else\(^{#1}\)\fi}
\begin{tabular}{lccc}
\toprule
            &\multicolumn{1}{c}{(1)}&\multicolumn{1}{c}{(2)}&\multicolumn{1}{c}{(3)}\\
            &\multicolumn{1}{c}{Click\ensuremath{\sb{t+1}}}&\multicolumn{1}{c}{Report\ensuremath{\sb{t+1}}}&\multicolumn{1}{c}{Safe handling\ensuremath{\sb{t+1}}}\\
\midrule
Click\ensuremath{\sb{t}}&       0.367\sym{***}&       0.076\sym{***}&       0.085\sym{***}\\
            &     (0.015)         &     (0.020)         &     (0.022)         \\\addlinespace
Constant    &      -1.089\sym{***}&      -2.340\sym{***}&      -2.332\sym{***}\\
            &     (0.154)         &     (0.354)         &     (0.419)         \\\addlinespace
\midrule
\textbf{APE}         &      \textbf{0.0705}         &     \textbf{ 0.0077 }        &     \textbf{ 0.0079}         \\
            &    (0.0032)         &    (0.0021)         &    (0.0021)         \\
Observations&     173,499         &     173,499         &     173,499         \\
Log pseudolikelihood&-52{,}210.089         &-31{,}480.284         &-28{,}893.343         \\
Pseudo \ensuremath{R\sp{2}}&       0.101         &       0.198         &       0.198         \\
\bottomrule
\addlinespace[0.2cm]
\end{tabular}
\end{singlespace}
\begin{flushleft}\footnotesize    \textit{Notes:} MSM+CRE pooled probit. CRE terms follow    Wooldridge~(2005): initial condition (\ensuremath{y\sb{i1}}),    Mundlak means of time-varying covariates, exposure count,    department group, and job status. Report and safe-handling models    additionally include initial reporting status. IPTW uses trimmed    stabilized weights. All models include next-campaign fixed effects.    Standard errors clustered by employee. APE row reports average    partial effect of \ensuremath{\text{Click}\sb{t}}.    \sym{*} \ensuremath{p<0.10}; \sym{**} \ensuremath{p<0.05};    \sym{***} \ensuremath{p<0.01}.    \end{flushleft}
\end{table}

A key implication of Table~\ref{tab:progression} is that persistence estimated without correcting for stable heterogeneity can be misleadingly interpreted as habit formation. To isolate this effect, we examine the within-person CRE estimate in Column~(3). Consistent with the bounding logic developed in Section \ref{subsec:combinedMSM.CRE}, leveraging the dynamic-panel structure to control for stable individual differences, this specification yields a conservative, lower-bound state-dependence effect. Because the model is nonlinear, we report and interpret this effect using the estimated average partial effect (APE) of lagged click behavior rather than the raw coefficient on $\text{Click}_t$. The resulting lower-bound APE is \creXTape\ (s.e. \creXTapeSE), corresponding to an increase of about 1.4 percentage points in the predicted probability of clicking at the next exposure. Relative to the MSM probit estimate in Column~(4), this implies that roughly 85--90\% of naively estimated persistence may be attributable to stable employee heterogeneity rather than true state dependence.\footnote{Calculated as the proportional reduction in the APE after accounting for stable heterogeneity: $(0.0975 - 0.0137)/0.0975 \approx 85.9\%$ relative to the MSM probit (Column 4), and $(0.1158 - 0.0137)/0.1158 \approx 88.2\%$ relative to the unadjusted probit (Column 2).} Taken together, the evidence supports a bounded interpretation: the true causal persistence, expressed as an APE, is plausibly between about 1.4 and \apeClickPP\ percentage points.

\subsection{Contextual transfer via cue similarity}

Table~\ref{tab:iptwcre_jaccard} estimates whether persistence depends on contextual overlap in phishing cues between the current and next campaign exposure. The interaction between prior clicking and cue similarity is positive for subsequent clicking (\jaccardIntCoeff), indicating that persistence strengthens when campaigns reuse similar persuasion cues.

\begin{table}[htbp]
\centering
\caption{Cue-similarity moderator (MSM+CRE)}
\label{tab:iptwcre_jaccard}
\begin{singlespace}
\def\sym#1{\ifmmode^{#1}\else\(^{#1}\)\fi}
\begin{tabular}{lccc}
\toprule
            &\multicolumn{1}{c}{(1)}&\multicolumn{1}{c}{(2)}&\multicolumn{1}{c}{(3)}\\
            &\multicolumn{1}{c}{Click\ensuremath{\sb{t+1}}}&\multicolumn{1}{c}{Report\ensuremath{\sb{t+1}}}&\multicolumn{1}{c}{Safe handling\ensuremath{\sb{t+1}}}\\
\midrule
Click\ensuremath{\sb{t}}&       0.281\sym{***}&       0.187\sym{***}&       0.212\sym{***}\\
            &     (0.026)         &     (0.040)         &     (0.042)         \\\addlinespace
Cue similarity (Jaccard)&       0.389\sym{***}&       0.190         &       0.274         \\
            &     (0.104)         &     (0.307)         &     (0.339)         \\\addlinespace
Click\ensuremath{\sb{t}} \ensuremath{\times} Cue similarity&       0.178\sym{***}&      -0.240\sym{***}&      -0.278\sym{***}\\
            &     (0.044)         &     (0.076)         &     (0.079)         \\\addlinespace
Constant    &      -1.458\sym{***}&      -2.397\sym{***}&      -2.437\sym{***}\\
            &     (0.173)         &     (0.501)         &     (0.600)         \\\addlinespace
\midrule
Observations&     173,499         &     173,499         &     173,499         \\
Log pseudolikelihood&-52{,}193.382         &-31{,}474.106         &-28{,}885.518         \\
Pseudo \ensuremath{R\sp{2}}&       0.101         &       0.198         &       0.198         \\
\bottomrule
\addlinespace[0.2cm]
\end{tabular}
\end{singlespace}
\begin{flushleft}\footnotesize    \textit{Notes:} MSM+CRE pooled probit. The Jaccard similarity index    measures the overlap in phishing-email tactic cues between the current    and next campaign exposure. A higher Jaccard value means the employee    encounters similar persuasion cues at \ensuremath{t+1}. All models    include CRE terms, next-campaign fixed effects, and trimmed stabilized    IPTW. Standard errors clustered by employee.    \sym{*} \ensuremath{p<0.10}; \sym{**} \ensuremath{p<0.05};    \sym{***} \ensuremath{p<0.01}.    \end{flushleft}
\end{table}
 
APEs show a steep gradient. When the next campaign shares no cues with the current campaign (Jaccard = 0), the effect of a prior click is about 4.1 percentage points. At Jaccard = 0.25 the effect is about 5.5 points, at the sample mean (0.475) it is about 7.0 points, at Jaccard = 0.75 it is about 9.0 points, and when cue sets are identical it reaches about 11.1 points. For reporting and safe handling, the click effect is largest when the next campaign is dissimilar (about 1.8 points at Jaccard = 0) and attenuates as similarity increases (approximately zero at high similarity). These patterns support cue-contingent persistence: vulnerability appears to persist primarily within cue families rather than as a uniform tendency to click.

\subsection{Teachable-moment design moderators}

Table~\ref{tab:iptwcre_design} tests whether teachable-moment page features moderate click persistence. The interaction terms indicate that annotated landing cues (\desLandCoeff), reporting pitch (\desPitchCoeff), and emotion or heuristic framing (\desEmotCoeff) attenuate persistence, whereas annotated-email cues modestly amplify persistence (\desEmailCoeff). The interaction terms are jointly significant (\ensuremath{\chi^2(4)} = \desJointChi).

\begin{table}[htbp]
\centering
\small
\caption{Teachable-moment design moderators (MSM+CRE)}
\label{tab:iptwcre_design}
\begin{singlespace}
\def\sym#1{\ifmmode^{#1}\else\(^{#1}\)\fi}
\begin{tabular}{lc}
\toprule
            &\multicolumn{1}{c}{(1)}\\
            &\multicolumn{1}{c}{Click\ensuremath{\sb{t+1}}}\\
\midrule
Click\ensuremath{\sb{t}}&       0.440\sym{***}\\
            &     (0.044)         \\\addlinespace
Annotated email&       0.446\sym{***}\\
            &     (0.108)         \\\addlinespace
Click\ensuremath{\sb{t}} \ensuremath{\times} Annotated email&       0.090\sym{**} \\
            &     (0.046)         \\\addlinespace
Annotated landing cues&       0.313\sym{**} \\
            &     (0.155)         \\\addlinespace
Click\ensuremath{\sb{t}} \ensuremath{\times} Annotated landing cues&      -0.122\sym{**} \\
            &     (0.050)         \\\addlinespace
Reporting pitch&       0.005         \\
            &     (0.149)         \\\addlinespace
Click\ensuremath{\sb{t}} \ensuremath{\times} Reporting pitch&      -0.170\sym{***}\\
            &     (0.033)         \\\addlinespace
Emotion or heuristic framing&      -0.392\sym{*}  \\
            &     (0.204)         \\\addlinespace
Click\ensuremath{\sb{t}} \ensuremath{\times} Emotion or heuristic framing&      -0.243\sym{***}\\
            &     (0.043)         \\\addlinespace
Constant    &      -0.984\sym{***}\\
            &     (0.167)         \\\addlinespace
\midrule
Observations&     173,499         \\
Log pseudolikelihood&-52{,}124.813         \\
Pseudo \ensuremath{R\sp{2}}&       0.102         \\
\bottomrule
\addlinespace[0.2cm]
\end{tabular}
\end{singlespace}
\begin{flushleft}\footnotesize    \textit{Notes:} MSM+CRE pooled probit. The dependent variable is    \ensuremath{\text{Click}\sb{t+1}}. Interaction terms capture how    education-page design elements moderate the persistence of a prior    click. Moderator main effects are included in the model but not    interpreted causally, because non-clickers do not view the education    page. Joint Wald test of the four interaction terms:    \ensuremath{\chi\sp{2}(4)} = 125.40,    \ensuremath{p} $< 0.001$.    All models include CRE terms, next-campaign FE, trimmed stabilized IPTW.    Standard errors clustered by employee.    \sym{*} \ensuremath{p<0.10}; \sym{**} \ensuremath{p<0.05};    \sym{***} \ensuremath{p<0.01}.    \end{flushleft}
\end{table}
 
The implied APEs highlight the managerial relevance of education-page design. With no design features, clicking at $t$ increases next-exposure clicking by about 6.2 percentage points. With emotion or heuristic framing alone, the effect falls to about 1.3 points (marginally significant). Combining emotion framing with a reporting pitch reduces the effect to about 0.1 points (not significant). With all features present, the effect is statistically indistinguishable from zero (about -0.1 points). In contrast, including the annotated-email feature weakly increases persistence, suggesting a potential counterproductive design element.

\subsection{Cue-specific persistence}

To assess whether persistence operates through distinct vulnerability channels, Table~\ref{tab:iptwcre_cues} estimates cue-specific click interactions. The one-at-a-time estimates show that transactional and urgency cues are associated with stronger persistence, while curiosity and authority cues are associated with weaker persistence.

\begin{table}[htbp]
\centering
\footnotesize
\caption{Cue-specific click persistence (MSM+CRE)}
\label{tab:iptwcre_cues}
\begin{singlespace}
\def\sym#1{\ifmmode^{#1}\else\(^{#1}\)\fi}
\renewcommand{\arraystretch}{0.95}\begin{tabular}{lcccccc}
\toprule
            &\multicolumn{1}{c}{(1)}&\multicolumn{1}{c}{(2)}&\multicolumn{1}{c}{(3)}&\multicolumn{1}{c}{(4)}&\multicolumn{1}{c}{(5)}&\multicolumn{1}{c}{(6)}\\
            &\multicolumn{1}{c}{Authority}&\multicolumn{1}{c}{Urgency}&\multicolumn{1}{c}{Financial}&\multicolumn{1}{c}{Curiosity}&\multicolumn{1}{c}{Internal}&\multicolumn{1}{c}{Transactional}\\
\midrule
Click\ensuremath{\sb{t}}&       0.415\sym{***}&       0.310\sym{***}&       0.358\sym{***}&       0.517\sym{***}&       0.392\sym{***}&       0.348\sym{***}\\
            &     (0.021)         &     (0.022)         &     (0.018)         &     (0.025)         &     (0.020)         &     (0.017)         \\\addlinespace
Click\ensuremath{\sb{t}} \ensuremath{\times} Authority&      -0.077\sym{***}&                     &                     &                     &                     &                     \\
            &     (0.027)         &                     &                     &                     &                     &                     \\\addlinespace
Click\ensuremath{\sb{t}} \ensuremath{\times} Urgency&                     &       0.101\sym{***}&                     &                     &                     &                     \\
            &                     &     (0.028)         &                     &                     &                     &                     \\\addlinespace
Click\ensuremath{\sb{t}} \ensuremath{\times} Financial&                     &                     &       0.027         &                     &                     &                     \\
            &                     &                     &     (0.029)         &                     &                     &                     \\\addlinespace
Click\ensuremath{\sb{t}} \ensuremath{\times} Curiosity&                     &                     &                     &      -0.207\sym{***}&                     &                     \\
            &                     &                     &                     &     (0.029)         &                     &                     \\\addlinespace
Click\ensuremath{\sb{t}} \ensuremath{\times} Internal&                     &                     &                     &                     &      -0.048\sym{*}  &                     \\
            &                     &                     &                     &                     &     (0.027)         &                     \\\addlinespace
Click\ensuremath{\sb{t}} \ensuremath{\times} Transactional&                     &                     &                     &                     &                     &       0.083\sym{***}\\
            &                     &                     &                     &                     &                     &     (0.031)         \\\addlinespace
Constant    &      -1.116\sym{***}&      -1.016\sym{***}&      -0.782\sym{***}&      -0.985\sym{***}&      -1.039\sym{***}&      -1.054\sym{***}\\
            &     (0.155)         &     (0.155)         &     (0.186)         &     (0.160)         &     (0.158)         &     (0.160)         \\\addlinespace
\midrule
Observations&     173,499         &     173,499         &     173,499         &     173,499         &     173,499         &     173,499         \\
Log pseudolikelihood&-52{,}204.329         &-52{,}200.681         &-52{,}204.244         &-52{,}181.122         &-52{,}201.651         &-52{,}204.000         \\
Pseudo \ensuremath{R\sp{2}}&       0.101         &       0.101         &       0.101         &       0.101         &       0.101         &       0.101         \\
\bottomrule
\addlinespace[0.2cm]
\end{tabular}
\end{singlespace}
\begin{flushleft}\footnotesize    \textit{Notes:} Each column reports a separate MSM+CRE pooled probit    where \ensuremath{\text{Click}\sb{t}} is interacted with a single    phishing-email tactic cue present in the campaign at \ensuremath{t}.    A positive interaction indicates that falling for an email with that cue    increases persistence (emboldening); a negative interaction indicates    attenuation. All models include CRE terms, next-campaign FE, and    trimmed stabilized IPTW. Standard errors clustered by employee.    \sym{*} \ensuremath{p<0.10}; \sym{**} \ensuremath{p<0.05};    \sym{***} \ensuremath{p<0.01}.    \end{flushleft}
\end{table}
 
Results from a joint model that includes all cue interactions simultaneously reinforce this pattern. To assess cue heterogeneity, we examine the coefficients on the cue-by-lagged-click interactions, which indicate whether a given cue amplifies or attenuates the persistence APE (the interaction coefficients themselves are in log-odds units, whereas we interpret persistence in probability units via APEs). For brevity, we do not tabulate the full set of interaction coefficients; instead, we report the joint Wald test and highlight the largest interactions. The strongest emboldening effects arise for transactional templates (\cueJointTransTemp) and urgency (\cueJointUrg), while curiosity strongly attenuates persistence (\cueJointCur) and authority weakly attenuates it (\cueJointAuth). A joint Wald test rejects equality of the cue interactions (\ensuremath{\chi^2(7)} = \cueJointChi, \ensuremath{p<0.001}). These findings are consistent with cue-specific learning: employees who fall for particular social-engineering tactics are more likely to repeat that mistake when similar cues recur.

\begin{table}[htbp]
\centering
\small
\caption{Cue type \ensuremath{\times} emotion framing interaction (MSM+CRE)}
\label{tab:iptwcre_cue_educ}
\begin{singlespace}
\def\sym#1{\ifmmode^{#1}\else\(^{#1}\)\fi}
\renewcommand{\arraystretch}{0.95}\begin{tabular}{lcccc}
\toprule
            &\multicolumn{1}{c}{(1)}&\multicolumn{1}{c}{(2)}&\multicolumn{1}{c}{(3)}&\multicolumn{1}{c}{(4)}\\
            &\multicolumn{1}{c}{Authority}&\multicolumn{1}{c}{Financial}&\multicolumn{1}{c}{Curiosity}&\multicolumn{1}{c}{Internal}\\
\midrule
Click\ensuremath{\sb{t}}&       0.457\sym{***}&       0.478\sym{***}&       0.517\sym{***}&       0.415\sym{***}\\
            &     (0.023)         &     (0.021)         &     (0.025)         &     (0.020)         \\\addlinespace
Click\ensuremath{\sb{t}} \ensuremath{\times} Authority&      -0.006         &                     &                     &                     \\
            &     (0.031)         &                     &                     &                     \\\addlinespace
Click\ensuremath{\sb{t}} \ensuremath{\times} Financial&                     &      -0.062\sym{*}  &                     &                     \\
            &                     &     (0.032)         &                     &                     \\\addlinespace
Click\ensuremath{\sb{t}} \ensuremath{\times} Curiosity&                     &                     &      -0.103\sym{***}&                     \\
            &                     &                     &     (0.032)         &                     \\\addlinespace
Click\ensuremath{\sb{t}} \ensuremath{\times} Internal&                     &                     &                     &       0.100\sym{***}\\
            &                     &                     &                     &     (0.032)         \\\addlinespace
Click\ensuremath{\sb{t}} \ensuremath{\times} Emotion framing&      -0.311\sym{***}&      -0.326\sym{***}&      -0.272\sym{***}&      -0.382\sym{***}\\
            &     (0.070)         &     (0.036)         &     (0.034)         &     (0.099)         \\\addlinespace
Click\ensuremath{\sb{t}} \ensuremath{\times} Auth \ensuremath{\times} Emotion&      -0.001         &                     &                     &                     \\
            &     (0.079)         &                     &                     &                     \\\addlinespace
Click\ensuremath{\sb{t}} \ensuremath{\times} Fin \ensuremath{\times} Emotion&                     &      -0.057         &                     &                     \\
            &                     &     (0.106)         &                     &                     \\\addlinespace
Click\ensuremath{\sb{t}} \ensuremath{\times} Intr \ensuremath{\times} Emotion&                     &                     &                     &       0.021         \\
            &                     &                     &                     &     (0.106)         \\\addlinespace
Constant    &      -1.145\sym{***}&      -0.764\sym{***}&      -1.050\sym{***}&      -1.053\sym{***}\\
            &     (0.156)         &     (0.188)         &     (0.161)         &     (0.159)         \\\addlinespace
\midrule
Observations&     173,499         &     173,499         &     173,499         &     173,499         \\
Log pseudolikelihood&-52{,}149.048         &-52{,}141.059         &-52{,}142.959         &-52{,}137.607         \\
Pseudo \ensuremath{R\sp{2}}&       0.102         &       0.102         &       0.102         &       0.102         \\
\bottomrule
\addlinespace[0.2cm]
\end{tabular}
\end{singlespace}
\begin{flushleft}\footnotesize    \textit{Notes:} Each column reports a separate MSM+CRE pooled probit    with a three-way interaction: \ensuremath{\text{Click}\sb{t} \times    \text{Cue} \times \text{Emotion framing}}. The three-way term tests    whether emotion/heuristic framing on the education page differentially    attenuates persistence depending on the type of cue in the phishing    email. The three-way interaction for the curiosity cue (Column 3) is omitted due to perfect collinearity; in our campaign configuration, emotion/heuristic framing was only ever deployed in scenarios that also featured curiosity cues.  All models include CRE terms, next-campaign FE, and trimmed    stabilized IPTW. Standard errors clustered by employee.    \sym{*} \ensuremath{p<0.10}; \sym{**} \ensuremath{p<0.05};    \sym{***} \ensuremath{p<0.01}.    \end{flushleft}
\end{table}
 
\subsection{Cue type \ensuremath{\times} education design}

Table~\ref{tab:iptwcre_cue_educ} examines whether emotion or heuristic framing is equally effective across phishing cues by estimating three-way interactions. Marginal effects indicate that emotion framing is a near-universal antidote, but its impact varies by cue family. For internal-context emails, persistence is largest without emotion framing: the marginal effect is about 13.0 percentage points, falling to about 2.6 percentage points when emotion framing is present.\footnote{Marginal effects are computed via the delta method; delta-method standard errors use the robust variance estimator.} For non-internal emails, emotion framing reduces persistence from about 7.6 percentage points to about 0.3 percentage points (not significant). For financial cues, emotion framing reduces persistence from about 7.9 percentage points to about 0.2 percentage points (not significant). These patterns suggest that internal-context phishing is particularly emboldening in the absence of psychologically salient education.

\subsection{Engagement decomposition}

Finally, Table~\ref{tab:iptwcre_engage} decomposes clicking at $t$ by engagement with the education page. Disengaged clickers (\ensuremath{\leq}10 seconds on the education page or timed out) exhibit an APE of \engDisengApe\ (\engDisengApePct\ percentage points). Engaged clickers (20--290 seconds) exhibit a larger APE of \engEngApe\ (\engEngApePct\ points). The difference is statistically significant (\ensuremath{\chi^2(1)} = \engWaldChi, \ensuremath{p} = \engWaldP). This pattern is consistent with an emboldening mechanism in which consequence-free engagement with training can reinforce a sense that clicking is safe, because the employee experiences a benign, friendly educational page rather than a negative operational consequence. We unpack the managerial implications of this ``phishing-test paradox'' further in Section \ref{subsec:emboldeningMechanism}.

\begin{table}[htbp]
\centering
\caption{Engagement decomposition (MSM+CRE)}
\label{tab:iptwcre_engage}
\begin{singlespace}
\def\sym#1{\ifmmode^{#1}\else\(^{#1}\)\fi}
\begin{tabular}{lc}
\toprule
            &\multicolumn{1}{c}{(1)}\\
            &\multicolumn{1}{c}{Click\ensuremath{\sb{t+1}}}\\
\midrule
Disengaged click\ensuremath{\sb{t}}&       0.422\sym{***}\\
            &     (0.019)         \\\addlinespace
Engaged click\ensuremath{\sb{t}}&       0.490\sym{***}\\
            &     (0.021)         \\\addlinespace
Constant    &      -1.103\sym{***}\\
            &     (0.163)         \\\addlinespace
\midrule
Observations&     167,155         \\
Log pseudolikelihood&-49{,}673.981         \\
Pseudo \ensuremath{R\sp{2}}&       0.102         \\
\bottomrule
\addlinespace[0.2cm]
\end{tabular}
\end{singlespace}
\begin{flushleft}\footnotesize    \textit{Notes:} MSM+CRE pooled probit. Clicking at \ensuremath{t}    is decomposed into disengaged clicks (education page viewed    \ensuremath{\leq}10 seconds or timed out at 300 seconds) and engaged    clicks (20--290 seconds on education page). The reference category is    not clicking. The estimation sample ($N=167,155$) reflects the deliberate exclusion of 6,344 buffer observations (education page view times of 11–19 seconds, or exceeding 290 seconds) to establish a sharp behavioral contrast between the engaged and disengaged cohorts.  Wald test of equality:    \ensuremath{\chi\sp{2}(1)} = 7.93,    \ensuremath{p} = 0.0049.    CRE terms include initial-condition analogs for each engagement type.    All models include next-campaign FE and trimmed stabilized IPTW.    Standard errors clustered by employee.    \sym{*} \ensuremath{p<0.10}; \sym{**} \ensuremath{p<0.05};    \sym{***} \ensuremath{p<0.01}.    \end{flushleft}
\end{table}
 
\subsection{Results Summary}
This study revisits a central premise of simulated phishing programs: that click-triggered feedback not only improves awareness but also reduces future susceptibility. The application of our analytical framework on a real-world dataset challenges the prevailing notion and reveals a more nuanced result. The  primary estimates from our analytical framework indicate that clicking at $t$ increases subsequent clicking by about \apeClickPP\ percentage points, even after jointly adjusting for time-varying confounding and stable employee heterogeneity. At the same time, reporting and safe handling increase by about \apeReportPP\ and \apeSafePP\ points, indicating that teachable moments can improve protective responses even when susceptibility persistence remains.

\section{Discussion and Implications}
A core theme of this \textit{Research Note} is that naive analyses of data collected through simulated phishing campaigns can systematically mislead. Click-triggered feedback in simulated phishing campaigns creates a dynamic causal setting in which two biases, time-varying confounding from endogenous training assignment and the conflation of true state dependence with stable employee heterogeneity, operate simultaneously and in opposite directions. Our framework addresses both, and the analytical progression makes their respective contributions transparent: conventional estimates are likely to overstate the role of habit formation because most observed persistence reflects who employees are, not what recent clicks have done to them. Importantly, the framework requires no data beyond what standard simulated phishing platforms already record, making it directly portable to other organizational settings. Below, we interpret the substantive findings through four lenses: information foraging and cue-contingent persistence, the emboldening mechanism, cue-specific vulnerability channels, and managerial implications, each with distinct consequences for how organizations design and evaluate their phishing programs.

\subsection{Information foraging and cue-contingent persistence}

The empirical patterns we report are consistent with an information-foraging interpretation of routine email triage. Employees face high message volume and must allocate limited attention across competing tasks. In such environments, behavior is often guided by the perceived information scent of an email, inferred from surface cues such as urgency language, transactional templates, and internal-context references \parencite{Pirolli1999InformationForaging, Pirolli2007InformationForaging}. When a particular cue combination has previously led an employee to click, encountering similar cues in a later campaign can reactivate the same heuristic, explaining why persistence is strongest when successive campaigns share cue categories. 

The same lens helps interpret why psychologically salient education can attenuate persistence. Education that explicitly articulates why the employee was susceptible and provides a simple reporting script can reduce the cognitive cost of a protective response under time pressure. In contrast, minimal or generic education may fail to change the underlying triage rule, leaving employees vulnerable to the same cue-triggered errors and potentially reinforcing a sense that clicking is consequence-free.

A key methodological implication is that longitudinal persistence is not synonymous with habit formation. The analytical progression and the within-person CRE estimate indicate that most persistence estimated in conventional models reflects stable heterogeneity rather than true state dependence. For research, this clarifies why prior studies that track repeat clickers may conflate persistent vulnerability with causal reinforcement. For practice, it suggests that ``repeat clicker'' lists primarily identify a stable high-risk subgroup rather than employees who are developing bad habits from recent failures. The distinction has direct resource implications: organizations that treat repeat clickers as candidates for benign training may be targeting a population that instead needs sustained technical safeguards or workflow redesign. 

\subsection{The emboldening mechanism and the phishing-test paradox}
\label{subsec:emboldeningMechanism}

Our results are consistent with an emboldening mechanism in consequence-free environments. In this setting, employees who click are redirected to a friendly education page, and clicking carries no penalty or immediate cost. Persistence is largest when education is minimal (no design features) and is essentially eliminated only when education includes psychologically salient framing and a concrete reporting script. The engagement decomposition strengthens this interpretation: employees who spend time on the education page nevertheless exhibit greater persistence than those who dismiss it. 

Together, these findings imply a \textit{phishing-test paradox}: simulated campaigns can inadvertently reinforce risky behavior unless the accompanying education is designed to create meaningful psychological cost and actionable defensive routines. The paradox is important to address in practice because it implies that simply increasing the frequency of simulated phishing tests, a common organizational response to high click rates, may backfire if the post-click education remains generic and consequence-free. More testing, in this case, can compound the problem rather than resolve it.  Indeed, this phenomenon aligns with the operational realities observed at our research site; anecdotally, the focal organization's Chief Information Security Officer (CISO) and security team advocated for stiffer penalties for persistent clickers in an effort to finally break these ingrained behaviors.

\subsection{Cue-specific vulnerability channels}

The cue-specific results suggest that phishing susceptibility is a bundle of distinct vulnerabilities rather than a single trait. Transactional and urgency cues are the most emboldening, consistent with the intuition that routine templates and time pressure encourage fast heuristic processing. Curiosity and authority cues attenuate persistence, suggesting that some cues are more self-correcting, perhaps because they are easier to recognize upon reflection or because their plausibility is more sensitive to organizational context. For organizations, this implies that training and simulation design should be cue targeted and should monitor which cue families generate persistent vulnerabilities.

\subsection{Managerial implications}

Three actionable implications emerge from our findings. First, evaluation of phishing programs should distinguish state dependence from stable heterogeneity, because the two imply fundamentally different intervention strategies. Our analytical framework provides a robust approach for making this distinction using data from standard phishing simulations. 

Second, education-page design is not a cosmetic choice: emotion or heuristic framing and an explicit reporting pitch can eliminate persistence, while some features may be counterproductive. Organizations should audit their post-click education pages with the same rigor they apply to phishing simulations themselves. 

Third, cue similarity and cue family should guide campaign sequencing. Reusing the same cue family can concentrate risk among vulnerable employees and may require stronger education, escalation, or complementary technical safeguards. Varying the cue mix across successive campaigns is a low-cost design lever that can reduce the accumulation of cue-specific vulnerability. 

\section{Limitations and Future Research}

This study has several limitations that future research can address. First, the correlated random effects approach relies on a finite Mundlak projection and initial-condition terms to proxy for stable employee heterogeneity. This projection is necessarily approximate and cannot capture all dimensions of unobserved vulnerability. For this reason, we interpret the within-person CRE estimate and the MSM+CRE estimate as informative bounds on true state dependence rather than point estimates. Alternate approaches, such as instrumental variable strategies or designs that exogenously vary training exposure, could sharpen this bound, but they require institutional variation in data that our research setting did not provide. 

Second, some cue \ensuremath{\times} education \ensuremath{\times} clicking cells in our configuration are sparse even with 17 campaigns, which can widen confidence intervals for higher-order interactions. Replication in settings with more campaigns, richer variation in education content, and alternative consequence structures would help establish the generality of these findings. In particular, organizations that impose real consequences for clicking (e.g., manager notification, penalties, costly training) would provide a valuable contrast to the consequence-free environment examined in our research. 

Third, the setting of our research is a large university, and the operational details of simulated phishing programs vary across organizations. How persistence and emboldening depend on governance choices such as consequences for clicking, the cadence of campaigns, and the integration of technical controls remains an open question. Our analytic framework is designed to be portable to different organizational scenarios, as it uses standard data simulated phishing platforms routinely collect. Whether the magnitudes of effects we report from our illustrative analysis generalize to various corporate or government settings is an empirical question that warrants further investigation. 

Finally, our coding of teachable-moment page features is coarse to facilitate parsimoniuous illustration. We classified pages along four design dimensions, but we did not measure the full range of instructional variation (e.g., reading level, visual design quality, or other interactive elements) that are likely to influence learning effectiveness. Future work with access to richer education-page metadata or randomized page designs could identify more precisely which instructional ingredients drive the attenuation effects we have reported.

\section{Conclusion}

Simulated phishing campaigns are only as useful as the conclusions organizations draw from their data. Our analysis shows that tracking who clicks repeatedly in phishing simulations and inferring that training is or isn't working is challenging. The data structure itself conflates stable individual vulnerability with genuine behavioral change, and training assignment is endogenous to the outcomes it is meant to improve. The analytical framework we develop disentangles these confounds and is portable to any organization that utilizes standard phishing simulation platforms. The substantive lesson we illustrate by applying our framework to a real-world phishing simulation dataset is equally transferable: Persistence is not a monolith, and it depends on which persuasion cues employees encounter, how successive campaigns are sequenced, and whether the post-click education is designed to create lasting psychological salience rather than mere compliance. Organizations that test more without teaching better, and that evaluate their programs without accounting for the biases we identify, risk compounding the very problem they set out to solve.

\bibliographystyle{plainnat}
\bibliography{references_main}

\end{document}


\title{Online Appendix for ``Breaking Bad Email Habits''}
\author{Authors' names not included for peer review}
\date{}
\maketitle

\appendix

This Online Appendix provides supplementary material for the main manuscript. It includes (i) structured summaries of related literature, (ii) documentation of scenario coding procedures and qualitative scenario descriptions, (iii) full scenario similarity matrices used to construct cue-similarity measures, (iv) additional diagnostics and robustness checks, and (v) additional methodological detail on the MSM+CRE estimator and comparison MSM results.

\section{Extended literature review tables}
\label{app:sec:ec_lit}

The main paper provides a focused literature review to motivate the research questions and the causal design. Here we provide structured summaries of prior work to enhance transparency and to document the empirical and theoretical foundations of the study. Tables~\ref{tab:susceptibility}--\ref{tab:detection} organize prior studies into four themes: (1) user susceptibility factors, (2) phishing message characteristics, (3) awareness and training interventions, and (4) detection and adjacent security topics.

\scriptsize
\begin{longtable}{p{0.17\textwidth} p{0.20\textwidth} p{0.28\textwidth} p{0.30\textwidth}}
\caption{Thematic Review: User Susceptibility Factors} \label{tab:susceptibility} \\
\toprule
\textbf{Title, authors, citation} & \textbf{Research questions} & \textbf{Data and methods} & \textbf{Findings}\\
\midrule
\endfirsthead
\toprule
\textbf{Title, authors, citation} & \textbf{Research questions} & \textbf{Data and methods} & \textbf{Findings}\\
\midrule
\endhead
\textbf{The Influence of Experiential and Dispositional Factors in Phishing: An Empirical Investigation of the Deceived, \citep*{Wright2010Experiential}} & This study seeks to understand the behavioral factors that make online users susceptible to phishing attacks. Drawing from deception detection research, it poses the central question: How do the experiential characteristics (computer self-efficacy, web experience, security knowledge) and dispositional characteristics (trust, perceived risk, suspicion) of users affect their likelihood of being deceived by a phishing email? The research models and empirically tests the influence of these six factors on the success of a phishing attempt. & A nine-week field study was conducted with 299 undergraduate business students. Participants were issued a confidential super-secure code'' (SSC) for a course management system and received training on internet security. In the final week, a phishing email from a fictitious database administrator'' was sent, requesting them to reply with their SSC. Deception success was a binary variable, coded as successful if a student provided the correct SSC. Experiential and dispositional factors were measured via surveys and a security knowledge assessment. The hypotheses were tested simultaneously using Structural Equation Modeling (SEM). & Deception was successful on 32\% of participants. The results show that experiential factors are highly influential. Higher computer self-efficacy, greater web experience, and better security knowledge all significantly decreased the likelihood of being deceived. Of the dispositional factors, only higher suspicion of humanity was found to significantly decrease susceptibility. Disposition to trust and perceived risk were not significant predictors in this context. Overall, the tested factors explained 36.4\% of the variance in deception success, highlighting that experience, training, and a suspicious mindset are the most effective defenses against phishing. \\
\midrule
\textbf{Why do people get phished? Testing individual differences in phishing vulnerability within an integrated, information processing model, \citep*{Vishwanath2011WhyDoPeople}} & This study develops and tests an integrated information processing model to explain why people fall for phishing attacks. It asks how individual factors (e.g., involvement, email load, knowledge, self-efficacy) and information processing (attention to cues and elaboration) jointly influence phishing susceptibility, and whether users process phishing emails centrally or peripherally. & A field survey collected 325 completed responses from university students who had recently received one of two real-world phishing emails at their university. Of these, 321 respondents were the intended victims and were split for model refinement (N=161) and validation (N=160). Participants were shown one phishing email and surveyed on their likelihood of responding, attention to cues (source, grammar, urgency, subject), elaboration, and individual factors such as email load and self-efficacy. The integrated model was tested using Structural Equation Modeling (SEM). & The model explained nearly 50\% of the variance in phishing susceptibility. Results indicate largely peripheral processing, with reliance on simple cues. Attention to the email's source and grammar decreased susceptibility, while attention to the subject line and urgency cues increased it. Habitual media use and high email load were strong direct predictors of increased susceptibility, consistent with vulnerability driven by routine and inattentive responding to salient message triggers. \\
\midrule
\textbf{Overconfidence in Phishing Email Detection, \citep*{Wang2016Overconfidence}} & This study investigates overconfidence in phishing email detection, focusing on cognitive and motivational antecedents of retrospective overconfidence (confidence after judgment). It examines cognitive effort, attention allocation, perceived familiarity, self-efficacy, and dispositional optimism as predictors of overprecision and overestimation. & A survey experiment was conducted with 600 U.S. participants sourced via Qualtrics. Each subject was shown 16 email images, randomly drawn from a pool of 50 phishing and genuine business emails. For each email, participants judged legitimacy and rated confidence. Overconfidence was measured as overprecision (mean confidence minus mean accuracy) and overestimation (estimated accuracy minus actual accuracy). Cognitive effort was proxied by response time; motivational factors were measured via survey scales. Hypotheses were tested using OLS regression. & Overconfidence was prevalent, with 80\% of participants exhibiting overprecision. Cognitive effort (more time spent) significantly reduced overconfidence, whereas variability in attention allocation increased it. Dispositional optimism and perceived familiarity significantly increased overconfidence; perceived self-efficacy had only a marginal effect. Confidence was a poor predictor of actual detection accuracy, highlighting the risk of relying on subjective certainty in security decisions.  \\
\midrule
\textbf{Which phish get caught? An exploratory study of individuals' susceptibility to phishing, \citep*{Moody2017WhichPhish}} & The study identifies situational and personality factors that predict susceptibility. It tests how message source (known vs. unknown), link type (text vs. numeric), personality traits (e.g., curiosity, risk propensity), and internet experience factors (e.g., general usage, internet anxiety) influence clicking behavior. & An ``ethical phishing'' experiment used a 2x2 factorial design, manipulating message source and link type. Data were collected from 595 undergraduate students who completed a survey measuring 12 independent variables identified through prior work and Delphi studies. Two weeks later, participants received a simulated phishing email. Clicking was modeled using multiple logistic regression. & Emails from a known source, higher curiosity, and greater risk propensity significantly increased susceptibility. Contrary to expectations, greater general internet usage and higher internet anxiety increased susceptibility. Trust and distrust were not significant as main effects, but post hoc analyses suggested they can predict susceptibility in specific contexts (e.g., when the email source is known), underscoring conditional and interaction-driven effects. \\
\midrule
\textbf{An Examination of the Effect of Recent Phishing Encounters on Phishing Susceptibility, \citep*{Chen2020RecentEncounters}} & This research asks how a recent phishing encounter affects perceived susceptibility and whether effects vary across users. It distinguishes the detection process (difficulty) from the detection outcome (failure) and examines how each contributes to perceived susceptibility. It also tests moderation by past detection success and phishing-news desensitization. & A survey of college students asked respondents to recall their most recent phishing encounter. Measures captured perceived detection process difficulty and whether the outcome was a failure, along with antecedents including detection self-efficacy, message involvement, email source familiarity, and past victimization. The model was analyzed with PLS-SEM. Moderation by past success and desensitization was tested via interaction terms. & Detection process difficulty and detection outcome failure each increased perceived phishing susceptibility. A more difficult detection process also increased the likelihood of outcome failure. Detection self-efficacy reduced perceived process difficulty (and indirectly reduced outcome failure through difficulty), while message involvement increased process difficulty (and indirectly increased outcome failure through difficulty). Past victimization reduced both process difficulty and outcome failure, whereas email source familiarity was not a significant predictor. The effects of process difficulty and outcome failure on perceived susceptibility were stronger for users with greater past detection success and for those more desensitized to phishing news. \\
\midrule
\textbf{Characteristics that predict phishing susceptibility: a review, \citep*{tornblad2021characteristics}} & As a foundational step toward a holistic predictive model, this review identifies user characteristics that prior studies report as significant predictors of phishing susceptibility and synthesizes them into a structured resource for future work. & A systematic literature review searched multiple academic databases and supplemented retrieval with backward citation tracking. The authors synthesized statistically significant predictors reported across the collected studies. & The review catalogs 32 distinct characteristics and organizes them into seven categories, including personality traits, demographics, educational background, cybersecurity experience and beliefs, platform experience, email behaviors, and work commitment style. The authors emphasize inconsistent results for many predictors and conclude that multi-factor models are necessary because no single factor is a strong predictor across contexts. \\
\midrule
\textbf{Phishing Susceptibility in Context: A Multilevel Information Processing Perspective on Deception Detection, \citep*{Wright2023Context}} & This study investigates why employees remain vulnerable despite training by conceptualizing susceptibility as a multilevel failure of deception detection within work context. It tests whether network position and workgroup conditions shape clicking and credential submission. & A field study was conducted with 133 employees across 15 workgroups in a university finance division. Data combined a phishing simulation (four realistic phishing emails) with surveys. Dependent variables were clicking and credential submission. Independent variables captured task-network and IT-advice-network centrality, reliance on formal IT support, and workgroup-level time pressure and resilience. Nested data were analyzed using hierarchical random-intercept logistic regression. & Reliance on formal IT support and higher workgroup time pressure increased susceptibility. Higher centrality in the work-task network reduced susceptibility. Workgroup resilience showed an unexpected positive association with clicking (but not with credential submission), suggesting a distinction between operational resilience and vigilance toward phishing and highlighting the importance of multilevel context beyond individual IT knowledge. \\
\midrule
\textbf{How do technology use patterns influence phishing susceptibility? A two-wave study of the role of reformulated locus of control, \citep*{Ayaburi2023TechnologyUse}} & This study examines how internal control expressed through technology use patterns and external control expressed through situational email cues jointly influence susceptibility. Using reformulated LOC theory, it differentiates \textit{automatic} and \textit{routine} technology use and tests how situational cues moderate these effects. & A two-wave design was employed. Study 1 was a field survey of 432 respondents who had received a real-world phishing email, analyzed using SEM. Study 2 was a scenario-based factorial survey experiment with 355 participants that manipulated situational cues (e.g., sender cues, grammar, urgency) and tested moderation using multi-group SEM. & Automatic technology use increased phishing susceptibility, while routine technology use decreased susceptibility in Study 1 (and was not significant in Study 2). A key boundary condition emerged in Study 2: automatic technology use increased susceptibility primarily when sender-related situational cues were present, highlighting an interaction between user tendencies and specific message cues. \\
\midrule
\textbf{Unraveling the behavioral influence of social media on phishing susceptibility: A Personality-Habit-Information Processing model, \citep*{Frauenstein2023PersonalityHabit}} & This study tests a Personality-Habit-Information Processing model in which Big Five traits shape risky social media habits, which in turn influence heuristic versus systematic processing and phishing susceptibility. It also examines social norms, computer self-efficacy, and perceived risk as additional predictors. & An online survey collected data from 215 social media users in South Africa. Measures included Big Five traits, social media habits, heuristic and systematic processing (elicited using multiple Facebook stimuli), social norms, computer self-efficacy, perceived risk, and phishing susceptibility (binary response to a simulated phishing email). Hypotheses were tested using SEM. & Extraversion, agreeableness, and neuroticism were positively linked to forming risky social media habits, whereas conscientiousness was negatively linked. Habits promoted heuristic processing and discouraged systematic processing. Habit and heuristic processing increased phishing susceptibility, while systematic processing decreased it. Social norms increased susceptibility and computer self-efficacy decreased it; perceived risk was not significant. \\ 
\midrule
\textbf{The Nexus of Mindfulness, Affect, and Information Processing in Phishing Identification: An Empirical Examination, \citep*{Bera2025MindfulnessNexus}} & This study examines how heuristic versus systematic processing influences phishing identification accuracy and how mindfulness (trait and domain-specific) and affective states shape processing routes and, ultimately, detection performance. & A scenario-based survey experiment retained 556 participants from a university and Amazon Mechanical Turk. Participants completed measures of trait and domain mindfulness, classified 10 emails (7 phishing, 3 legitimate), and then reported information processing tendencies and affective state. The model was analyzed using PLS-SEM. & Systematic processing improved phishing identification accuracy, while heuristic processing decreased it. Trait mindfulness had no significant direct effect on accuracy but improved accuracy indirectly by increasing domain mindfulness and promoting systematic processing. Positive affect was associated with lower detection accuracy, indicating that affective state can reduce evaluative quality even when participants report engagement in processing. \\ 
\midrule
\bottomrule
\end{longtable}
 
\begin{longtable}{p{0.17\textwidth} p{0.20\textwidth} p{0.28\textwidth} p{0.30\textwidth}}
\caption{Thematic Review: Phishing Message Characteristics} \label{tab:message} \\
\toprule
\textbf{Title, authors, citation} & \textbf{Research questions} & \textbf{Data and methods} & \textbf{Findings}\\
\midrule
\endfirsthead
\toprule
\textbf{Title, authors, citation} & \textbf{Research questions} & \textbf{Data and methods} & \textbf{Findings}\\
\midrule
\endhead
\textbf{Research Note---Influence Techniques in Phishing Attacks: An Examination of Vulnerability and Resistance, \citep*{Wright2014InfluenceTechniques}} & The study compares the effectiveness of different influence techniques used in phishing. It tests whether Cialdini’s six techniques change vulnerability and whether techniques requiring fictitious shared experience are less effective than self-determination-based techniques. & A large-scale field experiment involved 2,624 student participants from a Midwestern university. Participants were randomly assigned to one of 64 conditions and received a phishing email directing them to a fabricated university portal to enter credentials. Conditions embedded one or more of six influence techniques in a baseline message. Credential submission was analyzed using logistic regression, controlling for gender. & Four techniques significantly increased response likelihood (liking, scarcity, social proof, reciprocity), with liking showing the largest marginal effect. Consistency had no effect. Authority significantly decreased compliance. The results support the view that techniques offering higher self-determination can be more dangerous than low self-determination techniques such as authority. \\
\midrule
\textbf{Got Phished? Internet Security and Human Vulnerability, \citep*{Goel2017GotPhished}} & This study tests how contextualization, framing (gain vs. loss), and underlying motives influence susceptibility in a field phishing experiment. & A field experiment sent imitation phishing emails to 7,225 undergraduate students across eight conditions. Messages manipulated contextualization, framing, and motive. Outcomes were open and click-through rates; differences were analyzed using chi-square tests. & Contextualization was the most powerful factor. The course-registration threat produced a 37.3\% click rate, compared with the overall average click rate across conditions (13.3\%). Framing effects were mixed and depended on context. Messages appealing to acquisition motives were more effective than those appealing to social motives. Females were more likely to open emails, but not more likely to click. \\
\midrule
\textbf{Exploring susceptibility to phishing in the workplace, \citep*{WILLIAMS20181}} & Study 1 examines whether authority and urgency cues increase employee click behavior in workplace phishing simulations. Study 2 qualitatively explores how employees perceive susceptibility and workplace influences on behavior. & Mixed-methods. Study 1 analyzed historic data from nine simulated phishing emails sent to approximately 62,000 employees in a large UK public sector organization. Emails were coded for authority and urgency cues; logistic regression modeled their effects on clicking. Study 2 used six focus groups with 32 employees (N=32) from an international engineering firm and analyzed transcripts with a hybrid thematic approach. & Study 1 found that authority and urgency cues increased employee click rates, with authority exerting the stronger effect. Study 2 highlighted workplace factors such as routines, cognitive pressures, familiarity, and technical and social support. The study emphasizes a dissonance between effective influence cues (authority, urgency) and the cues employees report relying on (e.g., sender authenticity). \\
\midrule
\textbf{The influence of affective processing on phishing susceptibility, \citep*{Tian2024AffectiveProcessing}} & This study tests how three emotion dimensions (valence, arousal, certainty) affect phishing clicking, guided by the Affective Infusion Model. & A two-phase experiment validated manipulations in a pilot (N=241) and then ran a mock phishing experiment with 474 students using a 2x2x2 factorial design manipulating valence, arousal, and certainty. Clicking was modeled using logistic regression with controls. & Positive valence and low certainty increased clicking, whereas arousal did not have a significant effect. The results indicate that affective profiles combining positive valence with uncertainty can increase susceptibility, supporting an affective processing account of phishing behavior. \\
\bottomrule
\end{longtable}
 
\begin{longtable}{p{0.17\textwidth} p{0.20\textwidth} p{0.28\textwidth} p{0.30\textwidth}}
\caption{Thematic Review: Interventions and Training} \label{tab:interventions} \\
\toprule
\textbf{Title, authors, citation} & \textbf{Research questions} & \textbf{Data and methods} & \textbf{Findings}\\
\midrule
\endfirsthead
\toprule
\textbf{Title, authors, citation} & \textbf{Research questions} & \textbf{Data and methods} & \textbf{Findings}\\
\midrule
\endhead
\textbf{Phishing for phishing awareness, \citep*{Jansson2013Awareness}} & This study tests whether simulated phishing combined with embedded, on-demand training can cultivate resistance in a real-world setting without relying on volunteers. & A two-week field experiment at an institution in South Africa involved the full email user population (25,579). Across two weekly cycles, users received one of four simulated phishing emails. Users who reacted insecurely (clicked a link or opened an attachment) were shown an immediate warning screen and received a follow-up email with a link to optional training. Week-to-week changes in insecure reactions were used to infer learning. & Insecure reactions decreased by 42.63\% from week one to week two (after accounting for active users). A total of 976 users who reacted insecurely in the first week did not do so in the second, consistent with short-term learning from embedded feedback. The most enticing lure involved a pornographic scam. A limitation is that users may have warned one another during the exercise. \\
\midrule
\textbf{Training to Mitigate Phishing Attacks Using Mindfulness Techniques, \citep*{Jensen2017Mindfulness}} & This study evaluates whether mindfulness-based training improves resistance to phishing more than additional rule-based training, and whether training format (text-only vs. text-plus-graphics) changes effectiveness. & A field experiment involved 355 university students, faculty, and staff. Participants were assigned to mindfulness training or rule-based training (each delivered in text-only or text-plus-graphics format) or to a no-training control task (survey-only). Ten days later, a phishing simulation solicited university credentials. Outcomes were analyzed using logistic regression. & Mindfulness training reduced susceptibility relative to both rule-based training and the no-training control. Response rates were 7.5\% (mindfulness) versus 13.4\% (rule-based). Training format had no significant effect. The mindfulness approach was particularly effective for participants who were already confident or had low e-mail mindfulness. \\
\midrule
\textbf{Using susceptibility claims to motivate behaviour change in IT security, \citep*{Jensen2021Claims}} & Using Social Judgement Theory, this study investigates why susceptibility claims motivate behavior change for some threats but not others, focusing on overt (phishing) versus furtive (password cracking) threats. & A longitudinal field experiment with 138 university members manipulated high versus low susceptibility claims in a SETA intervention and measured both intentions and subsequent behaviors over 11 weeks. A pilot (N=136) validated manipulations. Data were analyzed using General Estimating Equations (GEE). & Susceptibility claims did not influence intentions for either phishing or password protection. However, susceptibility claims significantly motivated avoidance of overt attacks (phishing) but had no effect on precaution-taking for furtive attacks (password cracking), consistent with threat overtness shaping claim acceptance. \\
\midrule
\textbf{The Phishing Funnel Model: A Design Artifact to Predict User Susceptibility to Phishing Websites, \citep*{Abbasi2021PhishingFunnel}} & This study develops and evaluates the Phishing Funnel Model (PFM) to predict susceptibility over time and to guide interventions (warnings) that reduce risky interactions with phishing websites. & Two longitudinal field experiments in two organizations: a 12-month prediction study with 1,278 employees (49,373 interactions) and a 3-month intervention study with 1,218 employees testing PFM-guided warnings. The model uses support vector ordinal regression with a composite kernel (SVORCK) to predict the funnel stage reached. & PFM outperformed competing models, improving AUC by 8\%--52\% and achieving 96\% success in predicting visits to high-severity threats. In the intervention study, PFM-guided warnings reduced interactions across funnel stages, with 3--6$\times$ fewer transactions than comparison groups. A cost-benefit analysis projected savings of approximately \$1,960 per employee annually relative to competing approaches. \\
\midrule
\textbf{A comparison of features in a crowdsourced phishing warning system, \citep*{Nguyen2021Crowdsourced}} & This study tests which features of a crowdsourced warning system (number of reports, report source, accuracy rate, and disclosure of accuracy) influence detection accuracy, adherence, and anxiety. & A 2x2x2x2 between-subjects laboratory experiment with 438 undergraduate participants. Participants evaluated 10 emails in a simulated environment; warnings manipulated the four features. Outcomes included hits, false positives, adherence, and anxiety; data were analyzed using ANCOVA. & Warning accuracy rate was the most influential feature, improving hits, reducing false positives, increasing adherence, and lowering anxiety. Number of reports and report source showed limited main effects but mattered in interactions, indicating that warning effectiveness depends on both informational quality and perceived provenance. \\
\midrule
\textbf{Trust calibration of automated security IT artifacts: A multi-domain study of phishing-website detection tools, \citep*{Chen2021TrustTools}} & This study examines antecedents and consequences of calibrated trust in phishing-website detection tools using Automation Trust and Reliance (ATR) theory. & A full-factorial lab experiment with 865 participants manipulated detector accuracy, run-time, outcome severity, threat type (spoof vs. concocted), and domain (banking vs. pharmacy). Participants evaluated 10 websites using a custom tool. The model was tested using SEM group analysis. & Higher detector accuracy increased trust calibration, while greater outcome severity reduced it; run-time was not significant. Calibrated trust predicted reliance and future use intentions, and higher reliance improved user performance in avoiding phishing websites. \\
\midrule
\textbf{Improving Phishing Reporting Using Security Gamification, \citep*{Jensen2022Gamification}} & This study examines how gamification elements (feedback, incentives, and public vs. private leaderboards) affect reporting performance and motivation, and compares an optimized gamified system with training and warnings. & Three randomized lab experiments (total N=568) used a simulated email environment with reporting tasks and a prototype leaderboard system. Experiential and instrumental outcomes (hits and false positives) were analyzed using MANCOVA. & Gamification increased reporting. Public attribution combined with rewards and punishments best balanced reporting and accuracy. Rewards increased hits and false positives, while punishments decreased both. The optimized leaderboard outperformed traditional training and warnings but required careful calibration to manage false alarms. \\
\bottomrule
\end{longtable}
 
\begin{longtable}{p{0.17\textwidth} p{0.20\textwidth} p{0.28\textwidth} p{0.30\textwidth}}
\caption{Thematic Review: Phishing Detection and Adjacent Topics} \label{tab:detection} \\
\toprule
\textbf{Title, authors, citation} & \textbf{Research questions} & \textbf{Data and methods} & \textbf{Findings}\\
\midrule
\endfirsthead
\toprule
\textbf{Title, authors, citation} & \textbf{Research questions} & \textbf{Data and methods} & \textbf{Findings}\\
\midrule
\endhead
\textbf{Detecting Fake Websites: The Contribution of Statistical Learning Theory, \citep*{Abbasi2010FakeWebsites}} & This paper asks whether an SLT-based detection system using rich fraud cues and a custom kernel can outperform existing tools in detecting spoof and concocted fake websites. & Design science development of AZProtect using an SVM with a custom linear composite kernel and $\sim$6,000 fraud cues. Evaluated on 900 websites (200 legitimate, 350 concocted, 350 spoof) and compared to existing tools and alternative algorithms. & AZProtect achieved 92.56\% accuracy and outperformed benchmark tools by 10--15\%. Recall was high for spoof (96.29\%) and improved for concocted (86.57\%), where lookup-based tools were weak. The custom kernel materially improved performance relative to generic kernels. \\
\midrule
\textbf{Assessing the severity of phishing attacks: A hybrid data mining approach, \citep*{Chen2011HybridDataMining}} & This study asks whether hybrid data mining can assess phishing attack severity, operationalized as technical risk level and financial loss to targeted firms. & Hybrid mining of 1,030 phishing alerts from the Millersmiles database linked to CRSP firm financial data. Text mining extracted semantic features, combined with 25 financial variables. Models (Decision Tree, SVM, Neural Network) predicted risk level and financial impact categories. & Models predicted severity with up to 89\% accuracy. Combining textual alert features with firm financial data outperformed either source alone. Predictors differed by outcome: technical risk aligned with certain textual cues and firm size, while financial loss aligned with cues and firm financial exposure. \\ 
\midrule
\textbf{An Efficacious Method for Detecting Phishing Webpages through Target Domain Identification, \citep*{Ramesh2014TargetDomain}} & This research asks whether a method can both detect phishing webpages and identify their intended target domain without relying on blacklists or training data. & Proposes a Target Identification (TID) algorithm using the intersection of on-page domains and search-engine domains derived from extracted keywords. Candidate domains are ranked using inter-linkage structure, and DNS comparisons support classification. Evaluated on 4,574 live websites (1,200 legitimate, 3,374 phishing). & Achieved 99.62\% overall accuracy, 99.67\% true positive rate, and 0.5\% false positive rate. Target identification accuracy was 99.85\%. The approach supports detection and target identification for zero-day and embedded-object attacks. \\
\midrule
\textbf{Do phishing alerts impact global corporations? A firm value analysis, \citep*{Bose2014FirmValue}} & This study examines whether public phishing alerts produce negative stock-price reactions and abnormal trading volume and whether effects vary by firm, industry, country, and timing. & Event study of 1,942 phishing alerts targeting 259 firms across 32 countries (Jan 2003 to Dec 2007). Outcomes were cumulative abnormal returns (CAR) and cumulative abnormal volume (CAV) in a three-day event window. Moderation tested via subsampling and cross-sectional quantile regression. & Public phishing alerts produced statistically significant negative firm-value impacts, with average market capitalization losses of at least US\$411 million per alert. Abnormal trading volume also decreased. Effects were stronger for financial holding companies and for alerts released later in the sample period (2006--2007); US-listed firms were weakly more affected. \\
\midrule
\textbf{Enhancing Predictive Analytics for Anti-Phishing by Exploiting Website Genre Information, \citep*{Abbasi2015EnhancingPredictive}} & This study develops a genre tree kernel method to improve phishing detection accuracy and generalizability and evaluates its practical value for user decision-making. & Design science development of a genre tree kernel that represents website directories as trees labeled with 18 genres and uses an SVM classifier. Evaluated on 4,050 websites. A user study with 120 participants compared the method to AZProtect and a browser tool (IE Phishing Filter). & Achieved 97.01\% accuracy, with recall of 94.37\% (concocted) and 99.63\% (spoof), and improved generalizability across industries. In the user study, participants using the genre tree kernel exhibited lower phishing-site visitation rates (37\%--41\%) than benchmark tools (60\%--82\%) and were less willing to transact. \\
\midrule
\textbf{PhishWHO: Phishing Webpage Detection via Identity Keywords Extraction and Target Domain Name Finder, \citep*{Tan2016PhishWHO}} & This paper proposes PhishWHO, a technique that detects phishing by comparing a webpage’s claimed identity to its actual domain identity using identity keyword extraction and target-domain discovery. & A three-phase method extracts identity keywords using a weighted URL-token approach with an N-gram model, identifies the target domain via search-engine queries, and applies a three-tier matching system (including ccTLD variations and IP aliasing). & The method demonstrates strong performance in detecting phishing webpages and identifying the impersonated target domain, supporting detection beyond blacklist-based approaches by inferring identity from content and search results. \\
\midrule
\textbf{Getting phished on social media, \citep*{Vishwanath2017SocialMedia}} & This study examines how interface cues (friend count, profile photo) and device affordances shape susceptibility to a two-level social-network phishing attack on Facebook. & Field experiment with 127 undergraduate students manipulated phisher profile photo (present vs. absent) and friend count (0 vs. 10) for a level-1 friend request, followed by a level-2 information request two weeks later. Device use (smartphone vs. PC) was captured and analyzed with ordinal regression. & Friend count was the strongest heuristic, increasing acceptance odds substantially. Smartphone use increased susceptibility. Prior acceptance of the level-1 request strongly predicted compliance with the level-2 request, highlighting sequential vulnerability in social-network phishing. \\
\midrule
\textbf{United We Stand, Divided We Fall: An Autogenic Perspective on Empowering Cybersecurity in Organizations, \citep*{Durcikova2024Autogenic}} & This study asks how cybersecurity groups become (dis)empowered through emergent processes and interactions with organizational stakeholders. & Abductive multiple case study across 15 U.S. organizations with interviews (17 cybersecurity leaders and 34 additional informants) and archival data, combining qualitative coding with computational analysis of transcripts. & Empowerment emerges via a virtuous cycle in which bridging initiatives by security groups are met with stakeholder prioritizing, reinforcing protection outcomes; disempowerment emerges when bridging is met with thwarting and is reinforced by constraints and failures. \\
\midrule
\textbf{To alert or alleviate? A natural experiment on the effect of anti-phishing laws on corporate IT and security investments, \citep*{Wang2024Laws}} & This study tests whether state-level anti-phishing laws change corporate security and general IT investments and whether effects differ for single-state versus multi-state firms. & A difference-in-differences design analyzed a panel of $\sim$400,000 U.S. firms (2010--2017) using Harte-Hanks Market Intelligence. Michigan (2011) and Oregon (2015) served as treated states; propensity score matching improved comparability. & Anti-phishing laws increased security investments overall but reduced general IT investments on average. Multi-state firms increased both security and IT investments, while single-state firms reduced IT investments and showed a small, non-significant decrease in security investment. The study also reports spillover effects, with multi-state firms increasing investments across locations beyond the treated states. \\
\bottomrule
\end{longtable}
 
\normalsize

\clearpage

\section{Scenario coding procedures}
\label{app:sec:ec_coding}

This section provides additional detail on how scenario-level phishing cues and teachable-moment design elements were coded.

\subsection{Coding materials and unit of analysis}

Scenario coding was performed at the campaign template level. For each scenario, we used screenshots of (i) the phishing email as rendered to recipients and (ii) the post-click teachable-moment page displayed immediately after a click. Indicators were coded as binary (0/1) and were assigned conservatively: if a feature was not clearly present in the screenshots, it was coded as 0. Multiple characteristics could co-occur within a scenario.

\subsection{Email-side cues and teachable-moment design elements}

The email-side coding scheme captures five common social-engineering tactics: authority, urgency, financial appeal, curiosity, and internal context. In addition, we coded whether the email used a transactional template or an attachment lure. The teachable-moment coding captures whether the post-click page includes annotated email cues, annotated landing cues (including credential-entry cues), an explicit reporting pitch, and emotion or heuristic framing, as well as whether the page is scenario themed.

Operational definitions, the full codebook, and the scenario-level codes are reported below (Section~\ref{app:sec:ec_coding_tables}). Scenario themes and teachable-moment emphasis are reported in Section~\ref{app:sec:ec_themes}.

\clearpage

\subsection{Codebook, scenario-level codes, and qualitative themes}
\label{app:sec:ec_coding_tables}

\begin{table}[htbp]
\centering
\caption{Scenario-level coding scheme for phishing email cues and teachable-moment design}
\label{tab:codebook}
\begin{singlespace}
\begin{tabular}{p{3.2cm} p{11.2cm}}
\toprule
\textbf{Code} & \textbf{Operational definition (scenario-level; 0/1)}\\
\midrule
\multicolumn{2}{l}{\textit{Email tactic cues (used to compute cue similarity)}}\\
Authority (\texttt{auth}) & Message invokes hierarchy, official authority, or institutional power to compel action (e.g., compliance request from an administrator).\\
Urgency (\texttt{urg}) & Message pressures rapid action through deadlines, threats of loss, or time scarcity cues.\\
Financial (\texttt{fin}) & Message references payments, invoices, refunds, gift cards, payroll, or other monetary incentives or losses.\\
Curiosity (\texttt{cur}) & Message prompts exploration through uncertainty, intrigue, or novelty cues (e.g., new policy, unexpected document).\\
Internal context (\texttt{intr}) & Message is framed as originating from inside the organization or referencing internal processes, teams, or systems.\\
\addlinespace
\multicolumn{2}{l}{\textit{Teachable-moment page design features (moderators; shown only after a click)}}\\
Annotated email (\texttt{ann\_email}) & Educational page visually annotates the original email to highlight suspicious cues.\\
Annotated landing cues (\texttt{ann\_land}) & Educational page annotates or explains suspicious features of the landing page or credential-harvest request.\\
Reporting pitch (\texttt{report\_pitch}) & Educational page explicitly encourages reporting and provides procedural guidance (what to do and how).\\
Emotion or heuristic framing (\texttt{emot\_heur}) & Educational page uses affective cues (e.g., warning language) or simple heuristics/rules-of-thumb to support future detection.\\
Scenario-specific theme (\texttt{scen\_theme}) & Educational page content is tailored to the scenario narrative rather than generic guidance.\\
\bottomrule
\end{tabular}
\end{singlespace}
\begin{flushleft}\footnotesize
\textit{Notes:} Scenarios are campaign templates in the simulation platform. All codes are constant within a scenario and are merged to the employee-by-campaign panel. Cue similarity is computed as the Jaccard index over the five email tactic cues.
\end{flushleft}
\end{table}
 \begin{table}[htbp]
\centering
\caption{Scenario-level coded features used in similarity and design-moderation analyses}
\label{tab:scenario_codes}
\begin{singlespace}
\resizebox{\textwidth}{!}{\begin{tabular}{ccc ccccc ccccc}
\toprule
& & & \multicolumn{5}{c}{Email tactic cues} & \multicolumn{5}{c}{Teachable-moment page features} \\
\cmidrule(lr){4-8}\cmidrule(lr){9-13}
Campaign $t$ & Scenario & Start date & Auth & Urg & Fin & Cur & Intr & AnnEmail & AnnLand & ReportPitch & EmotHeur & Theme \\
\midrule
1 & 28 & 2016-06-07 & 0 & 0 & 1 & 1 & 0 & 0 & 0 & 0 & 0 & 0 \\
2 & 29 & 2016-07-11 & 0 & 0 & 1 & 1 & 0 & 0 & 0 & 0 & 0 & 0 \\
3 & 30 & 2016-08-02 & 1 & 1 & 0 & 0 & 1 & 1 & 0 & 0 & 0 & 0 \\
4 & 32 & 2016-08-17 & 1 & 1 & 0 & 0 & 1 & 1 & 0 & 0 & 0 & 0 \\
5 & 33 & 2016-11-15 & 1 & 1 & 0 & 0 & 1 & 0 & 1 & 0 & 0 & 0 \\
6 & 39 & 2017-08-01 & 0 & 0 & 1 & 1 & 0 & 1 & 0 & 0 & 0 & 0 \\
7 & 47 & 2018-02-14 & 1 & 1 & 1 & 0 & 0 & 1 & 0 & 0 & 0 & 1 \\
8 & 48 & 2018-03-15 & 0 & 1 & 0 & 1 & 0 & 1 & 0 & 0 & 0 & 0 \\
9 & 52 & 2018-06-12 & 1 & 1 & 0 & 1 & 0 & 1 & 0 & 1 & 0 & 0 \\
10 & 56 & 2018-11-13 & 1 & 1 & 0 & 1 & 1 & 1 & 1 & 0 & 1 & 0 \\
11 & 59 & 2019-01-30 & 1 & 0 & 1 & 0 & 1 & 1 & 0 & 1 & 0 & 0 \\
12 & 61 & 2019-05-15 & 0 & 0 & 1 & 1 & 0 & 1 & 0 & 1 & 0 & 0 \\
13 & 63 & 2019-06-12 & 0 & 0 & 0 & 1 & 1 & 1 & 0 & 1 & 1 & 0 \\
14 & 65 & 2019-07-24 & 1 & 0 & 0 & 1 & 1 & 1 & 1 & 0 & 1 & 0 \\
15 & 67 & 2019-10-02 & 1 & 1 & 0 & 1 & 1 & 1 & 0 & 1 & 1 & 0 \\
16 & 69 & 2019-12-11 & 0 & 0 & 1 & 1 & 0 & 1 & 0 & 1 & 1 & 0 \\
17 & 71 & 2020-02-11 & 0 & 1 & 0 & 1 & 1 & 1 & 0 & 1 & 1 & 0 \\
\bottomrule
\end{tabular}}
\end{singlespace}
\begin{flushleft}\footnotesize
\textit{Notes:} Codes are binary indicators (1=present, 0=absent) assigned at the campaign-scenario level. ``AnnLand'' denotes annotated landing page cues or credential-entry cues in the teachable-moment page. ``EmotHeur'' denotes affective framing and/or simple heuristics. See Table~\ref{tab:codebook} for operational definitions.
\end{flushleft}
\end{table}
 
\clearpage

\subsection{Scenario themes and teachable-moment emphasis}
\label{app:sec:ec_themes}

\vspace*{\fill}
\begin{center}
   For readability and formatting reasons, Table \ref{tab:scenario_themes_and_training} is presented in rotated form on the following page. Readers may proceed directly to the next page to view the complete table.
\end{center}
\vspace*{\fill}
\newpage

\begin{sidewaystable}[p]
\centering
\small
\caption{Phishing scenario themes, coded characteristics, and observed teachable-moment page features (scenarios with both phishing-email and education-page screenshots).}
\label{tab:scenario_themes_and_training}
\resizebox{\textwidth}{!}{\begin{tabular}{r p{6.6cm} p{5.0cm} p{4.6cm} p{7.3cm} p{7.3cm}}
\toprule
Scenario & Phish theme (email) & Phishing-email tactic cues & Email format & Teachable-moment page features & Notable education emphasis \\
\midrule
28 & ``Order Confirmation'' transactional email with ``Manage order'' button and order details & Financial=1, Curiosity=1 & Transactional template (button-style call to action) & Generic teachable-moment page (narrative guidance; no annotated screenshots) & General link-safety guidance (verify sender and link destination; avoid clicking unknown links) \\
29 & ``Order Confirmation'' transactional email with ``Manage order'' button and order details (same theme as Scenario 28) & Financial=1, Curiosity=1 & Transactional template (button-style call to action) & Generic teachable-moment page in ``quick tips'' format (bulleted; no annotated screenshots) & Concise rules of thumb (unknown sender caution; password-update and reset warnings) \\
30 & ``ACCOUNT ALERT'' email requesting an email-account update or validation via a link & Authority=1, Urgency=1, Internal=1 & Standard link email (non-transactional; no attachment lure) & Annotated phishing-email screenshot with numbered callouts and checklist-style guidance & Emphasizes internal-IT impersonation cues and ``do not validate credentials via email links'' \\
32 & ``URGENT'' internal-operations lure (printer queue or pending documents; login requested via link) & Authority=1, Urgency=1, Internal=1 & Standard link email (non-transactional; no attachment lure) & Annotated phishing-email screenshot with numbered callouts and checklist-style guidance & Highlights urgency framing, sender legitimacy checks, and caution with login links \\
33 & Account deletion or access-loss threat prompting verification on a credential-entry page & Authority=1, Urgency=1, Internal=1 & Standard link email (non-transactional; no attachment lure) & Annotated landing-page or credential-entry screenshot with callouts (URL and webpage legitimacy cues) & Focuses on recognizing fraudulent login pages (URL mismatch, webpage quality, credential-entry risks) \\
39 & ``Order Confirmation'' transactional email with order details and a prominent action button & Financial=1, Curiosity=1 & Transactional template (button-style call to action) & ``Can I click on the link?'' page with annotated phishing-email screenshot and callouts & Spotting transactional-template red flags (unexpected order, sender address, suspicious link destination) \\
47 & Tax-themed notice requesting verification of tax records via a link & Authority=1, Urgency=1, Financial=1 & Standard link email (non-transactional; no attachment lure) & Scenario-themed ``Tax Phishing'' page with annotated phishing-email screenshot & Tax-specific guidance and verification cues (unexpected tax notice, sender legitimacy, link inspection) \\
48 & Delivery-failure notice with link to a ``Delivery Label'' (parcel or courier lure) & Urgency=1, Curiosity=1 & Standard link email (non-transactional; no attachment lure) & ``Can I click on the link?'' page with annotated phishing-email screenshot and callouts & Emphasizes curiosity lures (unexpected package), generic greeting cues, and hover-to-inspect links \\
52 & Digital fax message waiting, with expiration framing and link to view message & Authority=1, Urgency=1, Curiosity=1 & Standard link email (non-transactional; no attachment lure) & Annotated phishing-email screenshot plus a distinct ``Report it'' section & Strong reporting encouragement coupled with concrete cue spotting (unexpected fax, link destination checks) \\
56 & ``RE: Meeting'' message-access lure (``Unable to share this email'') leading toward credential entry & Authority=1, Urgency=1, Curiosity=1, Internal=1 & Standard link email (non-transactional; no attachment lure) & Multi-stage education: annotated phishing email plus annotated credential-entry or landing page screenshot & Credential-harvest vigilance plus explicit discussion of persuasion triggers (curiosity and work-task pressure) \\
59 & ``Health Insurance'' portal link with generic greeting and visible spelling or quality cues & Authority=1, Financial=1, Internal=1 & Standard link email (non-transactional; no attachment lure) & Annotated phishing-email screenshot plus a distinct ``Report it'' section & Strong reporting norm plus concrete cue spotting (misspellings, sender legitimacy, portal-link caution) \\
61 & ``Electronic Invoice Notice'' with prominent ``Download Invoice'' call-to-action button & Financial=1, Curiosity=1 & Transactional template (button-style call to action) & Annotated phishing-email screenshot plus reporting-oriented guidance & Emphasizes ``unexpected invoice'' reasoning and link inspection (avoid opening unknown invoice links) \\
63 & ``Please review this file'' message with a spreadsheet attachment icon (document-review lure) & Curiosity=1, Internal=1 & Attachment lure (file-review prompt) & Annotated phishing-email screenshot plus a distinct ``Report it'' section & Frames file-review prompts as social-engineering tactics; emphasizes reporting and cautious handling of attachments \\
65 & ``Corporate social media portal'' invitation (join or signup) with link and credential-entry risk in the flow & Authority=1, Curiosity=1, Internal=1 & Standard link email (non-transactional; no attachment lure) & Annotated phishing email plus annotated login or landing page screenshot & Multi-stage training focused on credential-entry risks and psychological manipulation cues (curiosity and legitimacy heuristics) \\
67 & Office 365-style ``Incoming Emails Rejected'' notice with ``Retrieve messages'' link & Authority=1, Urgency=1, Curiosity=1, Internal=1 & Standard link email (non-transactional; no attachment lure) & Annotated phishing-email screenshot plus a distinct ``Report it'' section & Emphasizes internal-domain mismatch and reporting; highlights urgency cues as manipulation tactics \\
69 & ``Daily Deals'' promotional email emphasizing holiday deals with a ``View Deals Now'' button & Financial=1, Curiosity=1 & Transactional template (button-style call to action) & Annotated phishing-email screenshot plus a distinct ``Report it'' section & Emphasizes promotional-template cues (external sender, link destination checks) and highlights emotional triggers as attacker tactics; reinforces reporting norm \\
71 & ``Available?'' colleague-impersonation request for help with an ``urgent matter'' and a link for details (mobile signature) & Urgency=1, Curiosity=1, Internal=1 & Standard link email (non-transactional; no attachment lure) & Annotated phishing-email screenshot plus a distinct ``Report it'' section & Highlights sender-domain mismatch and social-engineering pressure (urgency and curiosity); emphasizes verifying sender identity and reporting suspected phishing \\
\bottomrule
\end{tabular}}
\end{sidewaystable}

\clearpage

\section{Scenario similarity matrices}
\label{app:sec:ec_similarity}

The cue-similarity extension in the main paper uses Jaccard similarity between the set of phishing-email tactic cues in scenario $t$ and the set of tactic cues in scenario $t+1$. This section reports full similarity matrices and descriptive summaries.

\subsection{Email-cue similarity}

Table~\ref{tab:ec_top_cue_pairs_jaccard} lists the most similar scenario pairs by email-cue Jaccard similarity. Table~\ref{tab:ec_jaccard_cues} reports the full Jaccard similarity matrix.

\begin{table}[htbp]
\centering
\caption{Most similar scenario pairs: email-cue Jaccard similarity (top 10).}
\label{tab:ec_top_cue_pairs_jaccard}
\begin{tabular}{rrrrr}
\toprule
Scenario $s$ & Scenario $s'$ & Similarity & Shared & Union \\
\midrule
56 & 67 & 1.00 & 4 & 4 \\
30 & 32 & 1.00 & 3 & 3 \\
30 & 33 & 1.00 & 3 & 3 \\
32 & 33 & 1.00 & 3 & 3 \\
28 & 29 & 1.00 & 2 & 2 \\
28 & 39 & 1.00 & 2 & 2 \\
28 & 61 & 1.00 & 2 & 2 \\
28 & 69 & 1.00 & 2 & 2 \\
29 & 39 & 1.00 & 2 & 2 \\
29 & 61 & 1.00 & 2 & 2 \\
\bottomrule
\end{tabular}
\end{table}
 \begin{sidewaystable}[p]
\centering\scriptsize
\caption{Jaccard similarity matrix for phishing-email tactic cues.}
\label{tab:ec_jaccard_cues}
\resizebox{\textwidth}{!}{\begin{tabular}{rrrrrrrrrrrrrrrrrr}
\toprule
Scenario & 28 & 29 & 30 & 32 & 33 & 39 & 47 & 48 & 52 & 56 & 59 & 61 & 63 & 65 & 67 & 69 & 71 \\
\midrule
28 & 1.00 & 1.00 & 0.00 & 0.00 & 0.00 & 1.00 & 0.25 & 0.33 & 0.25 & 0.20 & 0.25 & 1.00 & 0.33 & 0.25 & 0.20 & 1.00 & 0.25 \\
29 & 1.00 & 1.00 & 0.00 & 0.00 & 0.00 & 1.00 & 0.25 & 0.33 & 0.25 & 0.20 & 0.25 & 1.00 & 0.33 & 0.25 & 0.20 & 1.00 & 0.25 \\
30 & 0.00 & 0.00 & 1.00 & 1.00 & 1.00 & 0.00 & 0.50 & 0.25 & 0.50 & 0.75 & 0.50 & 0.00 & 0.25 & 0.50 & 0.75 & 0.00 & 0.50 \\
32 & 0.00 & 0.00 & 1.00 & 1.00 & 1.00 & 0.00 & 0.50 & 0.25 & 0.50 & 0.75 & 0.50 & 0.00 & 0.25 & 0.50 & 0.75 & 0.00 & 0.50 \\
33 & 0.00 & 0.00 & 1.00 & 1.00 & 1.00 & 0.00 & 0.50 & 0.25 & 0.50 & 0.75 & 0.50 & 0.00 & 0.25 & 0.50 & 0.75 & 0.00 & 0.50 \\
39 & 1.00 & 1.00 & 0.00 & 0.00 & 0.00 & 1.00 & 0.25 & 0.33 & 0.25 & 0.20 & 0.25 & 1.00 & 0.33 & 0.25 & 0.20 & 1.00 & 0.25 \\
47 & 0.25 & 0.25 & 0.50 & 0.50 & 0.50 & 0.25 & 1.00 & 0.25 & 0.50 & 0.40 & 0.50 & 0.25 & 0.00 & 0.20 & 0.40 & 0.25 & 0.20 \\
48 & 0.33 & 0.33 & 0.25 & 0.25 & 0.25 & 0.33 & 0.25 & 1.00 & 0.67 & 0.50 & 0.00 & 0.33 & 0.33 & 0.25 & 0.50 & 0.33 & 0.67 \\
52 & 0.25 & 0.25 & 0.50 & 0.50 & 0.50 & 0.25 & 0.50 & 0.67 & 1.00 & 0.75 & 0.20 & 0.25 & 0.25 & 0.50 & 0.75 & 0.25 & 0.50 \\
56 & 0.20 & 0.20 & 0.75 & 0.75 & 0.75 & 0.20 & 0.40 & 0.50 & 0.75 & 1.00 & 0.40 & 0.20 & 0.50 & 0.75 & 1.00 & 0.20 & 0.75 \\
59 & 0.25 & 0.25 & 0.50 & 0.50 & 0.50 & 0.25 & 0.50 & 0.00 & 0.20 & 0.40 & 1.00 & 0.25 & 0.25 & 0.50 & 0.40 & 0.25 & 0.20 \\
61 & 1.00 & 1.00 & 0.00 & 0.00 & 0.00 & 1.00 & 0.25 & 0.33 & 0.25 & 0.20 & 0.25 & 1.00 & 0.33 & 0.25 & 0.20 & 1.00 & 0.25 \\
63 & 0.33 & 0.33 & 0.25 & 0.25 & 0.25 & 0.33 & 0.00 & 0.33 & 0.25 & 0.50 & 0.25 & 0.33 & 1.00 & 0.67 & 0.50 & 0.33 & 0.67 \\
65 & 0.25 & 0.25 & 0.50 & 0.50 & 0.50 & 0.25 & 0.20 & 0.25 & 0.50 & 0.75 & 0.50 & 0.25 & 0.67 & 1.00 & 0.75 & 0.25 & 0.50 \\
67 & 0.20 & 0.20 & 0.75 & 0.75 & 0.75 & 0.20 & 0.40 & 0.50 & 0.75 & 1.00 & 0.40 & 0.20 & 0.50 & 0.75 & 1.00 & 0.20 & 0.75 \\
69 & 1.00 & 1.00 & 0.00 & 0.00 & 0.00 & 1.00 & 0.25 & 0.33 & 0.25 & 0.20 & 0.25 & 1.00 & 0.33 & 0.25 & 0.20 & 1.00 & 0.25 \\
71 & 0.25 & 0.25 & 0.50 & 0.50 & 0.50 & 0.25 & 0.20 & 0.67 & 0.50 & 0.75 & 0.20 & 0.25 & 0.67 & 0.50 & 0.75 & 0.25 & 1.00 \\
\bottomrule
\end{tabular}}
\end{sidewaystable}
 
\subsection{Teachable-moment feature similarity}

Because teachable-moment page features can also be similar across scenarios, Table~\ref{tab:ec_top_edu_pairs_jaccard} lists the most similar scenario pairs by education-feature Jaccard similarity and Table~\ref{tab:ec_jaccard_edu} reports the full matrix.

\begin{table}[htbp]
\centering
\caption{Most similar scenario pairs: education-feature Jaccard similarity (top 10).}
\label{tab:ec_top_edu_pairs_jaccard}
\begin{tabular}{rrrrr}
\toprule
Scenario $s$ & Scenario $s'$ & Similarity & Shared & Union \\
\midrule
56 & 65 & 1.00 & 3 & 3 \\
63 & 67 & 1.00 & 3 & 3 \\
63 & 69 & 1.00 & 3 & 3 \\
63 & 71 & 1.00 & 3 & 3 \\
67 & 69 & 1.00 & 3 & 3 \\
67 & 71 & 1.00 & 3 & 3 \\
69 & 71 & 1.00 & 3 & 3 \\
52 & 59 & 1.00 & 2 & 2 \\
52 & 61 & 1.00 & 2 & 2 \\
59 & 61 & 1.00 & 2 & 2 \\
\bottomrule
\end{tabular}
\end{table}
 \begin{sidewaystable}[p]
\centering\scriptsize
\caption{Jaccard similarity matrix for teachable-moment page features.}
\label{tab:ec_jaccard_edu}
\resizebox{\textwidth}{!}{\begin{tabular}{rrrrrrrrrrrrrrrrrr}
\toprule
Scenario & 28 & 29 & 30 & 32 & 33 & 39 & 47 & 48 & 52 & 56 & 59 & 61 & 63 & 65 & 67 & 69 & 71 \\
\midrule
28 & 1.00 & 1.00 & 0.00 & 0.00 & 0.00 & 0.00 & 0.00 & 0.00 & 0.00 & 0.00 & 0.00 & 0.00 & 0.00 & 0.00 & 0.00 & 0.00 & 0.00 \\
29 & 1.00 & 1.00 & 0.00 & 0.00 & 0.00 & 0.00 & 0.00 & 0.00 & 0.00 & 0.00 & 0.00 & 0.00 & 0.00 & 0.00 & 0.00 & 0.00 & 0.00 \\
30 & 0.00 & 0.00 & 1.00 & 1.00 & 0.00 & 1.00 & 0.50 & 1.00 & 0.50 & 0.33 & 0.50 & 0.50 & 0.33 & 0.33 & 0.33 & 0.33 & 0.33 \\
32 & 0.00 & 0.00 & 1.00 & 1.00 & 0.00 & 1.00 & 0.50 & 1.00 & 0.50 & 0.33 & 0.50 & 0.50 & 0.33 & 0.33 & 0.33 & 0.33 & 0.33 \\
33 & 0.00 & 0.00 & 0.00 & 0.00 & 1.00 & 0.00 & 0.00 & 0.00 & 0.00 & 0.33 & 0.00 & 0.00 & 0.00 & 0.33 & 0.00 & 0.00 & 0.00 \\
39 & 0.00 & 0.00 & 1.00 & 1.00 & 0.00 & 1.00 & 0.50 & 1.00 & 0.50 & 0.33 & 0.50 & 0.50 & 0.33 & 0.33 & 0.33 & 0.33 & 0.33 \\
47 & 0.00 & 0.00 & 0.50 & 0.50 & 0.00 & 0.50 & 1.00 & 0.50 & 0.33 & 0.25 & 0.33 & 0.33 & 0.25 & 0.25 & 0.25 & 0.25 & 0.25 \\
48 & 0.00 & 0.00 & 1.00 & 1.00 & 0.00 & 1.00 & 0.50 & 1.00 & 0.50 & 0.33 & 0.50 & 0.50 & 0.33 & 0.33 & 0.33 & 0.33 & 0.33 \\
52 & 0.00 & 0.00 & 0.50 & 0.50 & 0.00 & 0.50 & 0.33 & 0.50 & 1.00 & 0.25 & 1.00 & 1.00 & 0.67 & 0.25 & 0.67 & 0.67 & 0.67 \\
56 & 0.00 & 0.00 & 0.33 & 0.33 & 0.33 & 0.33 & 0.25 & 0.33 & 0.25 & 1.00 & 0.25 & 0.25 & 0.50 & 1.00 & 0.50 & 0.50 & 0.50 \\
59 & 0.00 & 0.00 & 0.50 & 0.50 & 0.00 & 0.50 & 0.33 & 0.50 & 1.00 & 0.25 & 1.00 & 1.00 & 0.67 & 0.25 & 0.67 & 0.67 & 0.67 \\
61 & 0.00 & 0.00 & 0.50 & 0.50 & 0.00 & 0.50 & 0.33 & 0.50 & 1.00 & 0.25 & 1.00 & 1.00 & 0.67 & 0.25 & 0.67 & 0.67 & 0.67 \\
63 & 0.00 & 0.00 & 0.33 & 0.33 & 0.00 & 0.33 & 0.25 & 0.33 & 0.67 & 0.50 & 0.67 & 0.67 & 1.00 & 0.50 & 1.00 & 1.00 & 1.00 \\
65 & 0.00 & 0.00 & 0.33 & 0.33 & 0.33 & 0.33 & 0.25 & 0.33 & 0.25 & 1.00 & 0.25 & 0.25 & 0.50 & 1.00 & 0.50 & 0.50 & 0.50 \\
67 & 0.00 & 0.00 & 0.33 & 0.33 & 0.00 & 0.33 & 0.25 & 0.33 & 0.67 & 0.50 & 0.67 & 0.67 & 1.00 & 0.50 & 1.00 & 1.00 & 1.00 \\
69 & 0.00 & 0.00 & 0.33 & 0.33 & 0.00 & 0.33 & 0.25 & 0.33 & 0.67 & 0.50 & 0.67 & 0.67 & 1.00 & 0.50 & 1.00 & 1.00 & 1.00 \\
71 & 0.00 & 0.00 & 0.33 & 0.33 & 0.00 & 0.33 & 0.25 & 0.33 & 0.67 & 0.50 & 0.67 & 0.67 & 1.00 & 0.50 & 1.00 & 1.00 & 1.00 \\
\bottomrule
\end{tabular}}
\end{sidewaystable}
 
\subsection{Alternative similarity metric: simple matching coefficient}

As a robustness-oriented descriptive measure, we also report a simple matching coefficient (SMC), defined as the fraction of the five tactic indicators that match (including joint absences). Tables~\ref{tab:ec_smc_cues} and \ref{tab:ec_smc_edu} report SMC matrices for phishing-email tactics and teachable-moment features, respectively.

\begin{table}[htbp]
\centering
\caption{Most similar scenario pairs: email-cue simple matching coefficient (top 10).}
\label{tab:ec_top_cue_pairs_smc}
\begin{tabular}{rrr}
\toprule
Scenario $s$ & Scenario $s'$ & SMC \\
\midrule
56 & 67 & 1.00 \\
30 & 32 & 1.00 \\
30 & 33 & 1.00 \\
32 & 33 & 1.00 \\
28 & 29 & 1.00 \\
28 & 39 & 1.00 \\
28 & 61 & 1.00 \\
28 & 69 & 1.00 \\
29 & 39 & 1.00 \\
29 & 61 & 1.00 \\
\bottomrule
\end{tabular}
\end{table}
 \begin{sidewaystable}[p]
\centering\scriptsize
\caption{Simple matching coefficient matrix for phishing-email tactic cues.}
\label{tab:ec_smc_cues}
\resizebox{\textwidth}{!}{\begin{tabular}{rrrrrrrrrrrrrrrrrr}
\toprule
Scenario & 28 & 29 & 30 & 32 & 33 & 39 & 47 & 48 & 52 & 56 & 59 & 61 & 63 & 65 & 67 & 69 & 71 \\
\midrule
28 & 1.00 & 1.00 & 0.00 & 0.00 & 0.00 & 1.00 & 0.40 & 0.60 & 0.40 & 0.20 & 0.40 & 1.00 & 0.60 & 0.40 & 0.20 & 1.00 & 0.40 \\
29 & 1.00 & 1.00 & 0.00 & 0.00 & 0.00 & 1.00 & 0.40 & 0.60 & 0.40 & 0.20 & 0.40 & 1.00 & 0.60 & 0.40 & 0.20 & 1.00 & 0.40 \\
30 & 0.00 & 0.00 & 1.00 & 1.00 & 1.00 & 0.00 & 0.60 & 0.40 & 0.60 & 0.80 & 0.60 & 0.00 & 0.40 & 0.60 & 0.80 & 0.00 & 0.60 \\
32 & 0.00 & 0.00 & 1.00 & 1.00 & 1.00 & 0.00 & 0.60 & 0.40 & 0.60 & 0.80 & 0.60 & 0.00 & 0.40 & 0.60 & 0.80 & 0.00 & 0.60 \\
33 & 0.00 & 0.00 & 1.00 & 1.00 & 1.00 & 0.00 & 0.60 & 0.40 & 0.60 & 0.80 & 0.60 & 0.00 & 0.40 & 0.60 & 0.80 & 0.00 & 0.60 \\
39 & 1.00 & 1.00 & 0.00 & 0.00 & 0.00 & 1.00 & 0.40 & 0.60 & 0.40 & 0.20 & 0.40 & 1.00 & 0.60 & 0.40 & 0.20 & 1.00 & 0.40 \\
47 & 0.40 & 0.40 & 0.60 & 0.60 & 0.60 & 0.40 & 1.00 & 0.40 & 0.60 & 0.40 & 0.60 & 0.40 & 0.00 & 0.20 & 0.40 & 0.40 & 0.20 \\
48 & 0.60 & 0.60 & 0.40 & 0.40 & 0.40 & 0.60 & 0.40 & 1.00 & 0.80 & 0.60 & 0.00 & 0.60 & 0.60 & 0.40 & 0.60 & 0.60 & 0.80 \\
52 & 0.40 & 0.40 & 0.60 & 0.60 & 0.60 & 0.40 & 0.60 & 0.80 & 1.00 & 0.80 & 0.20 & 0.40 & 0.40 & 0.60 & 0.80 & 0.40 & 0.60 \\
56 & 0.20 & 0.20 & 0.80 & 0.80 & 0.80 & 0.20 & 0.40 & 0.60 & 0.80 & 1.00 & 0.40 & 0.20 & 0.60 & 0.80 & 1.00 & 0.20 & 0.80 \\
59 & 0.40 & 0.40 & 0.60 & 0.60 & 0.60 & 0.40 & 0.60 & 0.00 & 0.20 & 0.40 & 1.00 & 0.40 & 0.40 & 0.60 & 0.40 & 0.40 & 0.20 \\
61 & 1.00 & 1.00 & 0.00 & 0.00 & 0.00 & 1.00 & 0.40 & 0.60 & 0.40 & 0.20 & 0.40 & 1.00 & 0.60 & 0.40 & 0.20 & 1.00 & 0.40 \\
63 & 0.60 & 0.60 & 0.40 & 0.40 & 0.40 & 0.60 & 0.00 & 0.60 & 0.40 & 0.60 & 0.40 & 0.60 & 1.00 & 0.80 & 0.60 & 0.60 & 0.80 \\
65 & 0.40 & 0.40 & 0.60 & 0.60 & 0.60 & 0.40 & 0.20 & 0.40 & 0.60 & 0.80 & 0.60 & 0.40 & 0.80 & 1.00 & 0.80 & 0.40 & 0.60 \\
67 & 0.20 & 0.20 & 0.80 & 0.80 & 0.80 & 0.20 & 0.40 & 0.60 & 0.80 & 1.00 & 0.40 & 0.20 & 0.60 & 0.80 & 1.00 & 0.20 & 0.80 \\
69 & 1.00 & 1.00 & 0.00 & 0.00 & 0.00 & 1.00 & 0.40 & 0.60 & 0.40 & 0.20 & 0.40 & 1.00 & 0.60 & 0.40 & 0.20 & 1.00 & 0.40 \\
71 & 0.40 & 0.40 & 0.60 & 0.60 & 0.60 & 0.40 & 0.20 & 0.80 & 0.60 & 0.80 & 0.20 & 0.40 & 0.80 & 0.60 & 0.80 & 0.40 & 1.00 \\
\bottomrule
\end{tabular}}
\end{sidewaystable}
 \begin{table}[htbp]
\centering
\caption{Most similar scenario pairs: education-feature simple matching coefficient (top 10).}
\label{tab:ec_top_edu_pairs_smc}
\begin{tabular}{rrr}
\toprule
Scenario $s$ & Scenario $s'$ & SMC \\
\midrule
56 & 65 & 1.00 \\
63 & 67 & 1.00 \\
63 & 69 & 1.00 \\
63 & 71 & 1.00 \\
67 & 69 & 1.00 \\
67 & 71 & 1.00 \\
69 & 71 & 1.00 \\
52 & 59 & 1.00 \\
52 & 61 & 1.00 \\
59 & 61 & 1.00 \\
\bottomrule
\end{tabular}
\end{table}
 \begin{sidewaystable}[p]
\centering\scriptsize
\caption{Simple matching coefficient matrix for teachable-moment page features.}
\label{tab:ec_smc_edu}
\resizebox{\textwidth}{!}{\begin{tabular}{rrrrrrrrrrrrrrrrrr}
\toprule
Scenario & 28 & 29 & 30 & 32 & 33 & 39 & 47 & 48 & 52 & 56 & 59 & 61 & 63 & 65 & 67 & 69 & 71 \\
\midrule
28 & 1.00 & 1.00 & 0.80 & 0.80 & 0.80 & 0.80 & 0.60 & 0.80 & 0.60 & 0.40 & 0.60 & 0.60 & 0.40 & 0.40 & 0.40 & 0.40 & 0.40 \\
29 & 1.00 & 1.00 & 0.80 & 0.80 & 0.80 & 0.80 & 0.60 & 0.80 & 0.60 & 0.40 & 0.60 & 0.60 & 0.40 & 0.40 & 0.40 & 0.40 & 0.40 \\
30 & 0.80 & 0.80 & 1.00 & 1.00 & 0.60 & 1.00 & 0.80 & 1.00 & 0.80 & 0.60 & 0.80 & 0.80 & 0.60 & 0.60 & 0.60 & 0.60 & 0.60 \\
32 & 0.80 & 0.80 & 1.00 & 1.00 & 0.60 & 1.00 & 0.80 & 1.00 & 0.80 & 0.60 & 0.80 & 0.80 & 0.60 & 0.60 & 0.60 & 0.60 & 0.60 \\
33 & 0.80 & 0.80 & 0.60 & 0.60 & 1.00 & 0.60 & 0.40 & 0.60 & 0.40 & 0.60 & 0.40 & 0.40 & 0.20 & 0.60 & 0.20 & 0.20 & 0.20 \\
39 & 0.80 & 0.80 & 1.00 & 1.00 & 0.60 & 1.00 & 0.80 & 1.00 & 0.80 & 0.60 & 0.80 & 0.80 & 0.60 & 0.60 & 0.60 & 0.60 & 0.60 \\
47 & 0.60 & 0.60 & 0.80 & 0.80 & 0.40 & 0.80 & 1.00 & 0.80 & 0.60 & 0.40 & 0.60 & 0.60 & 0.40 & 0.40 & 0.40 & 0.40 & 0.40 \\
48 & 0.80 & 0.80 & 1.00 & 1.00 & 0.60 & 1.00 & 0.80 & 1.00 & 0.80 & 0.60 & 0.80 & 0.80 & 0.60 & 0.60 & 0.60 & 0.60 & 0.60 \\
52 & 0.60 & 0.60 & 0.80 & 0.80 & 0.40 & 0.80 & 0.60 & 0.80 & 1.00 & 0.40 & 1.00 & 1.00 & 0.80 & 0.40 & 0.80 & 0.80 & 0.80 \\
56 & 0.40 & 0.40 & 0.60 & 0.60 & 0.60 & 0.60 & 0.40 & 0.60 & 0.40 & 1.00 & 0.40 & 0.40 & 0.60 & 1.00 & 0.60 & 0.60 & 0.60 \\
59 & 0.60 & 0.60 & 0.80 & 0.80 & 0.40 & 0.80 & 0.60 & 0.80 & 1.00 & 0.40 & 1.00 & 1.00 & 0.80 & 0.40 & 0.80 & 0.80 & 0.80 \\
61 & 0.60 & 0.60 & 0.80 & 0.80 & 0.40 & 0.80 & 0.60 & 0.80 & 1.00 & 0.40 & 1.00 & 1.00 & 0.80 & 0.40 & 0.80 & 0.80 & 0.80 \\
63 & 0.40 & 0.40 & 0.60 & 0.60 & 0.20 & 0.60 & 0.40 & 0.60 & 0.80 & 0.60 & 0.80 & 0.80 & 1.00 & 0.60 & 1.00 & 1.00 & 1.00 \\
65 & 0.40 & 0.40 & 0.60 & 0.60 & 0.60 & 0.60 & 0.40 & 0.60 & 0.40 & 1.00 & 0.40 & 0.40 & 0.60 & 1.00 & 0.60 & 0.60 & 0.60 \\
67 & 0.40 & 0.40 & 0.60 & 0.60 & 0.20 & 0.60 & 0.40 & 0.60 & 0.80 & 0.60 & 0.80 & 0.80 & 1.00 & 0.60 & 1.00 & 1.00 & 1.00 \\
69 & 0.40 & 0.40 & 0.60 & 0.60 & 0.20 & 0.60 & 0.40 & 0.60 & 0.80 & 0.60 & 0.80 & 0.80 & 1.00 & 0.60 & 1.00 & 1.00 & 1.00 \\
71 & 0.40 & 0.40 & 0.60 & 0.60 & 0.20 & 0.60 & 0.40 & 0.60 & 0.80 & 0.60 & 0.80 & 0.80 & 1.00 & 0.60 & 1.00 & 1.00 & 1.00 \\
\bottomrule
\end{tabular}}
\end{sidewaystable}

\clearpage

\section{Weight diagnostics}
\label{app:sec:weights}

\begin{figure}[htbp]
\centering
\includegraphics[width=0.85\textwidth]{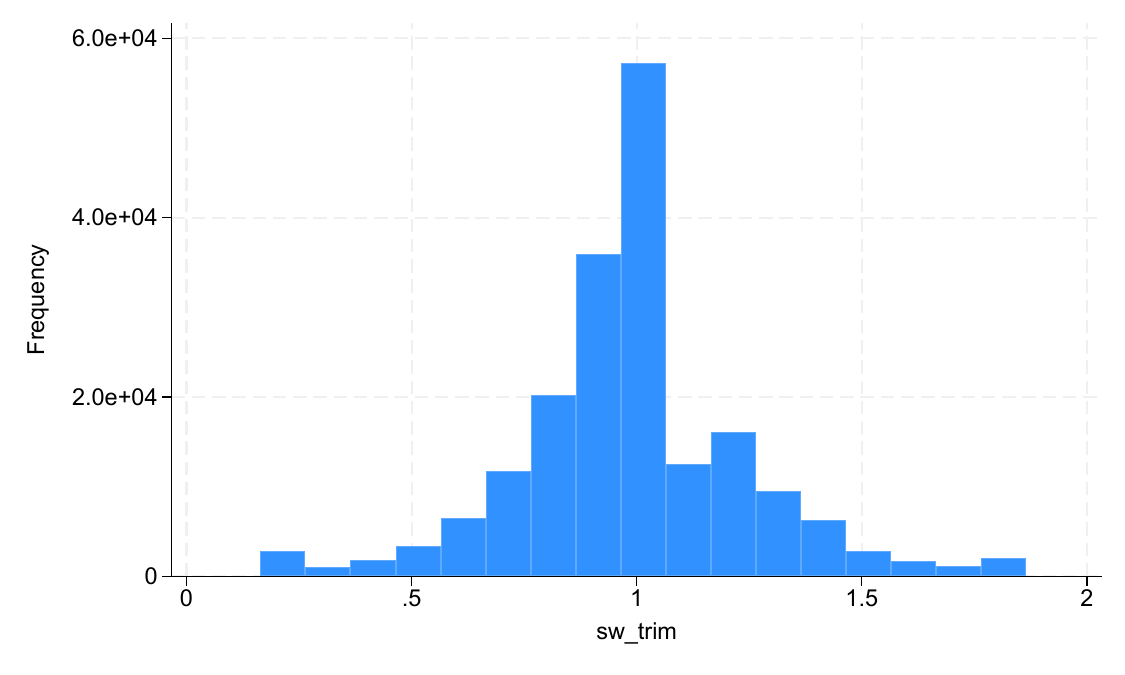}
\caption{Histogram of trimmed stabilized IPTW.}
\label{fig:ec_weights_hist}
\end{figure}

Table~\ref{tab:ecWeights} summarizes the stabilized IPTW distribution, and Figure~\ref{fig:ec_weights_hist} plots a histogram of trimmed stabilized weights.

\begin{table}[htbp]\centering
\caption{Stabilized inverse probability weights (IPTW) summary statistics.}\label{tab:ecWeights}
\begin{tabular}{lrrrrrrrrr}
\toprule
                    &        Mean&          SD&          P1&          P5&         P50&         P95&         P99&         Min&         Max\\
\hline
sw                  &       0.989&       0.270&       0.164&       0.575&       0.986&       1.425&       1.775&       0.000&       3.983\\
sw\_trim             &       0.987&       0.256&       0.164&       0.575&       0.986&       1.425&       1.775&       0.164&       1.775\\
\hline
Observations        &      192840&            &            &            &            &            &            &            &            \\
\bottomrule
\end{tabular}
\end{table}
 
\clearpage

\section{Dynamic panel identification: CRE and MSM+CRE}
\label{app:sec:ec_cre}

This section provides additional intuition for the correlated random effects (CRE) approach in dynamic nonlinear panels and for the combined MSM+CRE estimator used in the main paper.

\subsection{Why stable heterogeneity matters for persistence}

In repeated simulated-phishing settings, some employees are persistently more click prone for stable reasons such as job routines, technical environment, or enduring attention constraints. If these stable differences are omitted, longitudinal persistence estimates can overstate true state dependence.

\subsection{CRE for dynamic nonlinear panels}

We follow \textcite{Wooldridge2005InitialConditions}. The approach models the unobserved individual effect as correlated with observed covariates via a Mundlak projection \parencite{Mundlak1978Pooling}. Empirically, we include (i) an initial-condition indicator (click status at the first observed exposure), (ii) individual means of time-varying covariates used in the outcome models, (iii) exposure count, and (iv) department-group and job-status controls. These terms relax strict random-effects independence and reduce confounding of state dependence with stable heterogeneity.

\subsection{Combining IPTW with CRE}

IPTW targets time-varying confounding induced by click-triggered feedback. CRE targets stable heterogeneity. We combine the two by estimating pooled probit models with CRE terms using trimmed stabilized IPTW as probability weights and clustering standard errors by employee. A key implication is a bounding interpretation: a within-person CRE estimate provides a conservative lower bound on state dependence, while MSM+CRE provides an upper bound when the CRE projection is approximate.

\subsection{Analytical progression table}

\begin{table}[htbp]
\centering
\caption{Analytical progression: identifying click persistence}
\label{tab:progression}
\begin{singlespace}
\def\sym#1{\ifmmode^{#1}\else\(^{#1}\)\fi}
\begin{tabular}{lccccc}
\toprule
&\multicolumn{1}{c}{(1)}&\multicolumn{1}{c}{(2)}&\multicolumn{1}{c}{(3)}&\multicolumn{1}{c}{(4)}&\multicolumn{1}{c}{(5)}\\
&\multicolumn{1}{c}{FE-LPM}&\multicolumn{1}{c}{Probit}&\multicolumn{1}{c}{CRE probit}&\multicolumn{1}{c}{MSM probit}&\multicolumn{1}{c}{MSM+CRE}\\
\midrule
Click\ensuremath{\sb{t}}&      -0.088\sym{***}&      &        &    &        \\
      &     (0.003)      &   &   &   &   \\\addlinespace
Click\ensuremath{\sb{t}}&   &       0.550\sym{***}&       0.093\sym{***}&       0.475\sym{***}&       0.367\sym{***}\\
&   &     (0.014)   &     (0.015)   &     (0.014)   &     (0.015)   \\\addlinespace
\midrule
\textbf{APE}   &    --   &      \textbf{0.1158}   &      \textbf{0.0137}   &      \textbf{0.0975}   &      \textbf{0.0705}   \\
Addresses   & --  &     Neither   & Stable het.   &TV confounding   &        Both   \\
Observations&     173,499   &     173,499   &     173,499   &     173,499   &     173,499   \\
\bottomrule
\addlinespace[0.2cm]
\end{tabular}
\end{singlespace}
\begin{flushleft}\footnotesize    \textit{Notes:} All columns report the coefficient on    \ensuremath{\text{Click}\sb{t}} from the indicated estimator.    Column~(1) is a fixed-effects linear probability model (subject to    Nickell bias). Columns~(2)--(5) are probit models; APEs are reported    at the foot of the table. CRE terms include initial condition    (\ensuremath{y\sb{i1}}), individual means of time-varying covariates    (Mundlak terms), exposure count, department group, and job status.    IPTW uses stabilized inverse probability of treatment weights trimmed at    the 1st and 99th percentiles. All models include next-campaign fixed    effects. Standard errors are clustered by employee.    \sym{*} \ensuremath{p<0.10}; \sym{**} \ensuremath{p<0.05};    \sym{***} \ensuremath{p<0.01}.    \end{flushleft}
\end{table}
 
\clearpage

\section{Safe-handling Paradox}
\label{app:sec:ec_safe_paradox}

The main paper reports that clicking at exposure $t$ can increase both subsequent clicking and subsequent safe handling. These outcomes are not logical inverses because safe handling is not the complement of clicking. Instead, safe handling is a joint event that requires both not clicking and reporting at the next exposure.

\subsection{Definitions and the four behavioral states at the next exposure}

Let $A_{t+1}\in\{0,1\}$ denote clicking at the next exposure and $R_{t+1}\in\{0,1\}$ denote reporting at the next exposure. The joint state space contains four categories:

\begin{center}
\begin{tabular}{ll}
\toprule
State & Definition \\
\midrule
(A) & $A_{t+1}=1,\; R_{t+1}=1$ \\
(B) & $A_{t+1}=1,\; R_{t+1}=0$ \\
(C) & $A_{t+1}=0,\; R_{t+1}=0$ \\
(D) & $A_{t+1}=0,\; R_{t+1}=1$ \quad (\textbf{safe handling})\\
\bottomrule
\end{tabular}
\end{center}

The relevant marginal probabilities are:
\begin{align}
\Pr(A_{t+1}=1) &= \Pr(A)+\Pr(B),\\
\Pr(A_{t+1}=0) &= \Pr(C)+\Pr(D)=1-\Pr(A_{t+1}=1),\\
\Pr( \text{Safe}_{t+1}=1) &= \Pr(D)=\Pr(A_{t+1}=0,\;R_{t+1}=1).
\end{align}

Not clicking is the complement of clicking, but safe handling is a subset of not clicking. Not clicking includes both safe handling (D) and inaction (C).

\subsection{Illustrative example}

Consider two counterfactual worlds for exposure $t$:

\begin{itemize}
\item \textbf{Control world:} the employee does not click at $t$ ($A_t=0$), so the click-triggered feedback page is not shown.
\item \textbf{Treated world:} the employee clicks at $t$ ($A_t=1$), which triggers the teachable-moment feedback page.
\end{itemize}

Assume we have 100 otherwise comparable employee-exposure instances in each world, and we tabulate outcomes at the next exposure $t+1$.

\begin{table}[htbp]
\centering
\caption{Illustrative joint outcomes at $t+1$ for 100 control versus 100 treated exposure instances.}
\begin{tabular}{lcc}
\toprule
Joint state at $t+1$ & Control world ($A_t=0$) & Treated world ($A_t=1$)\\
\midrule
(A) $A_{t+1}=1,\;R_{t+1}=1$ & 1 & 1 \\
(B) $A_{t+1}=1,\;R_{t+1}=0$ & 8 & 9 \\
(D) $A_{t+1}=0,\;R_{t+1}=1$ \quad (\textbf{safe}) & 6 & 8 \\
(C) $A_{t+1}=0,\;R_{t+1}=0$ & 85 & 82 \\
\midrule
Total & 100 & 100 \\
\bottomrule
\end{tabular}
\end{table}

In this example, clicking at $t$ increases subsequent clicking (9\% to 10\%) and also increases safe handling (6\% to 8\%). This can occur because safe handling depends on both the share of non-click exposures and the propensity to report among non-clickers. Specifically,
\begin{equation}
\Pr(A_{t+1}=0,\;R_{t+1}=1)=\Pr(R_{t+1}=1\mid A_{t+1}=0)\Pr(A_{t+1}=0).
\label{eq:ec_safe_product}
\end{equation}
A treatment can reduce $\Pr(A_{t+1}=0)$ (by increasing clicking) and still increase safe handling if it increases $\Pr(R_{t+1}=1\mid A_{t+1}=0)$ sufficiently.

\clearpage

\section{MSM-only comparison tables}
\label{app:sec:ec_msm_compare}

For transparency, this section reports MSM (IPTW logit) estimates that adjust time-varying confounding but do not correct for stable employee heterogeneity. These tables correspond to earlier versions of the analysis and are provided as comparison benchmarks.  In the main body of the \textit{Research Note}, we used an outcome model with a probit link function for comparability with other models. 

\begin{table}[htbp]
\centering
\caption{Marginal structural model (MSM) estimates without stable-heterogeneity correction}
\label{tab:mainMSM}
\begin{singlespace}
\def\sym#1{\ifmmode^{#1}\else\(^{#1}\)\fi}
\begin{tabular}{lccc}
\toprule
            &\multicolumn{1}{c}{(1)}&\multicolumn{1}{c}{(2)}&\multicolumn{1}{c}{(3)}\\
            &\multicolumn{1}{c}{Click\ensuremath{\sb{t+1}}}&\multicolumn{1}{c}{Report\ensuremath{\sb{t+1}}}&\multicolumn{1}{c}{Safe handling\ensuremath{\sb{t+1}}}\\
\midrule
Click\ensuremath{\sb{t}}&       0.894\sym{***}&       0.232\sym{***}&       0.239\sym{***}\\
            &     (0.026)         &     (0.038)         &     (0.041)         \\\addlinespace
Constant    &      -1.726\sym{***}&      -4.120\sym{***}&      -4.231\sym{***}\\
            &     (0.044)         &     (0.125)         &     (0.131)         \\\addlinespace
\midrule
Observations&     173,499         &     173,499         &     173,499         \\
Log pseudolikelihood&-53{,}431.263         &-35{,}278.016         &-32{,}254.573         \\
Pseudo \ensuremath{R\sp{2}}&       0.080         &       0.101         &       0.104         \\
\bottomrule
\end{tabular}
\end{singlespace}
\begin{flushleft}\footnotesize    \textit{Notes:} These MSM estimates adjust for time-varying confounding but do not separate true state dependence from stable individual heterogeneity; we report them in the eCompanion as a comparison benchmark.  Each column reports a weighted pooled-logit marginal structural    model with stabilized inverse probability of treatment weights (IPTW).    The treatment is \ensuremath{\text{Click}\_{t}}, which also triggers exposure    to the platform's teachable-moment educational page. Outcomes are measured at    the employee's next observed campaign exposure. All models include next-campaign    fixed effects (indicators for the campaign at \ensuremath{t+1}) to absorb    systematic differences in subsequent campaign difficulty. Stabilized weights are    trimmed at the 1st and 99th percentiles. Standard errors are clustered by employee.    \sym{*} \ensuremath{p<0.10}; \sym{**} \ensuremath{p<0.05};    \sym{***} \ensuremath{p<0.01}.    \end{flushleft}
\end{table}
 
\begin{table}[htbp]
\centering
\footnotesize
\caption{MSM extensions (without stable-heterogeneity correction): cue similarity and design moderators}
\label{tab:extensions}
\begin{singlespace}
\def\sym#1{\ifmmode^{#1}\else\(^{#1}\)\fi}
\begin{tabular}{lcccc}
\toprule
                    &\multicolumn{1}{c}{(1)}&\multicolumn{1}{c}{(2)}&\multicolumn{1}{c}{(3)}&\multicolumn{1}{c}{(4)}\\
                    &\multicolumn{1}{c}{Click\ensuremath{\sb{t+1}}}&\multicolumn{1}{c}{Report\ensuremath{\sb{t+1}}}&\multicolumn{1}{c}{Click\ensuremath{\sb{t+1}}}&\multicolumn{1}{c}{Report\ensuremath{\sb{t+1}}}\\
\midrule
Click\ensuremath{\sb{t}}&       0.795\sym{***}&       0.437\sym{***}&       1.199\sym{***}&       0.619\sym{***}\\
                    &     (0.048)         &     (0.084)         &     (0.083)         &     (0.197)         \\\addlinespace
Cue similarity (Jaccard)&       0.716\sym{***}&       1.041\sym{***}&                     &                     \\
                    &     (0.185)         &     (0.401)         &                     &                     \\\addlinespace
Click\ensuremath{\sb{t}} \ensuremath{\times} Cue similarity&       0.197\sym{**} &      -0.454\sym{***}&                     &                     \\
                    &     (0.081)         &     (0.165)         &                     &                     \\\addlinespace
Annotated email     &                     &                     &       0.776\sym{***}&       1.151\sym{*}  \\
                    &                     &                     &     (0.192)         &     (0.621)         \\\addlinespace
Click\ensuremath{\sb{t}} \ensuremath{\times} Annotated email&                     &                     &      -0.018         &      -0.763\sym{***}\\
                    &                     &                     &     (0.085)         &     (0.190)         \\\addlinespace
Annotated landing cues&                     &                     &       0.670\sym{**} &       0.977         \\
                    &                     &                     &     (0.279)         &     (0.634)         \\\addlinespace
Click\ensuremath{\sb{t}} \ensuremath{\times} Annotated landing cues&                     &                     &      -0.297\sym{***}&       0.346\sym{**} \\
                    &                     &                     &     (0.090)         &     (0.161)         \\\addlinespace
Reporting pitch     &                     &                     &      -0.031         &       0.616         \\
                    &                     &                     &     (0.310)         &     (0.522)         \\\addlinespace
Click\ensuremath{\sb{t}} \ensuremath{\times} Reporting pitch&                     &                     &      -0.338\sym{***}&       0.744\sym{***}\\
                    &                     &                     &     (0.059)         &     (0.103)         \\\addlinespace
Emotion or heuristic framing&                     &                     &      -0.682\sym{*}  &      -0.000         \\
                    &                     &                     &     (0.410)         &     (0.401)         \\\addlinespace
Click\ensuremath{\sb{t}} \ensuremath{\times} Emotion or heuristic framing&                     &                     &      -0.439\sym{***}&      -0.458\sym{***}\\
                    &                     &                     &     (0.081)         &     (0.115)         \\\addlinespace
Constant            &      -2.540\sym{***}&      -4.911\sym{***}&      -2.031\sym{***}&      -4.506\sym{***}\\
                    &     (0.192)         &     (0.421)         &     (0.091)         &     (0.230)         \\\addlinespace
\midrule
Observations        &     173,499         &     173,499         &     173,499         &     173,499         \\
Log pseudolikelihood&-53{,}419.746         &-35{,}270.335         &-53{,}333.220         &-35{,}209.602         \\
Pseudo \ensuremath{R\sp{2}}&       0.080         &       0.101         &       0.081         &       0.103         \\
\bottomrule
\end{tabular}
\end{singlespace}
\begin{flushleft}\footnotesize    \textit{Notes:} These are MSM (IPTW logit) estimates reported as a comparison benchmark; they adjust for time-varying confounding but do not correct for stable employee heterogeneity.  Columns (1) and (2) test contextual transfer by interacting    \ensuremath{\text{Click}\_{t}} with the Jaccard similarity of phishing-email    tactic cues between campaigns \ensuremath{t} and \ensuremath{t+1}. Columns (3)    and (4) test teachable-moment page design moderators (coded at the scenario level    for campaign \ensuremath{t}). Moderator main effects are not interpreted causally    because non-clickers do not view the teachable-moment page; interaction terms    capture how design elements change the effect of a prior click (and exposure to    the teachable moment) on subsequent behavior. All models include next-campaign    fixed effects, use trimmed stabilized IPTW, and cluster standard errors by employee.    \sym{*} \ensuremath{p<0.10}; \sym{**} \ensuremath{p<0.05};    \sym{***} \ensuremath{p<0.01}.    \end{flushleft}
\end{table}
 
\begin{table}[htbp]
\centering
\footnotesize
\caption{MSM additional analyses (without stable-heterogeneity correction): heterogeneity and decay in click persistence}
\label{tab:heterogeneity}
\begin{singlespace}
\def\sym#1{\ifmmode^{#1}\else\(^{#1}\)\fi}
\renewcommand{\arraystretch}{0.95}\begin{tabular}{lcccc}
\toprule
                    &\multicolumn{1}{c}{(1)}&\multicolumn{1}{c}{(2)}&\multicolumn{1}{c}{(3)}&\multicolumn{1}{c}{(4)}\\
                    &\multicolumn{1}{c}{Role}&\multicolumn{1}{c}{Tenure}&\multicolumn{1}{c}{Prior click history}&\multicolumn{1}{c}{Time gap}\\
\midrule
Click\ensuremath{\sb{t}}&       0.969\sym{***}&       0.626\sym{***}&       1.244\sym{***}&       1.293\sym{***}\\
                    &     (0.049)         &     (0.046)         &     (0.038)         &     (0.145)         \\\addlinespace
Click\ensuremath{\sb{t}} \ensuremath{\times} Staff&      -0.145\sym{***}&                     &                     &                     \\
                    &     (0.056)         &                     &                     &                     \\\addlinespace
Click\ensuremath{\sb{t}} \ensuremath{\times} Unknown role&      -0.043         &                     &                     &                     \\
                    &     (0.065)         &                     &                     &                     \\\addlinespace
Click\ensuremath{\sb{t}} \ensuremath{\times} Tenure Q2&                     &       0.247\sym{***}&                     &                     \\
                    &                     &     (0.069)         &                     &                     \\\addlinespace
Click\ensuremath{\sb{t}} \ensuremath{\times} Tenure Q3&                     &       0.323\sym{***}&                     &                     \\
                    &                     &     (0.072)         &                     &                     \\\addlinespace
Click\ensuremath{\sb{t}} \ensuremath{\times} Tenure Q4&                     &       0.364\sym{***}&                     &                     \\
                    &                     &     (0.071)         &                     &                     \\\addlinespace
Click\ensuremath{\sb{t}} \ensuremath{\times} Tenure missing&                     &       0.301\sym{***}&                     &                     \\
                    &                     &     (0.065)         &                     &                     \\\addlinespace
Click\ensuremath{\sb{t}} \ensuremath{\times} Prior clicks = 1&                     &                     &      -0.530\sym{***}&                     \\
                    &                     &                     &     (0.056)         &                     \\\addlinespace
Click\ensuremath{\sb{t}} \ensuremath{\times} Prior clicks = 2&                     &                     &      -0.750\sym{***}&                     \\
                    &                     &                     &     (0.070)         &                     \\\addlinespace
Click\ensuremath{\sb{t}} \ensuremath{\times} Prior clicks \ensuremath{\geq} 3&                     &                     &      -0.837\sym{***}&                     \\
                    &                     &                     &     (0.078)         &                     \\\addlinespace
Log days to next exposure&                     &                     &                     &      -0.071\sym{**} \\
                    &                     &                     &                     &     (0.028)         \\\addlinespace
Click\ensuremath{\sb{t}} \ensuremath{\times} Log days to next exposure&                     &                     &                     &      -0.089\sym{***}\\
                    &                     &                     &                     &     (0.032)         \\\addlinespace
Constant            &      -1.986\sym{***}&      -1.399\sym{***}&      -2.075\sym{***}&      -1.553\sym{***}\\
                    &     (0.049)         &     (0.050)         &     (0.052)         &     (0.116)         \\\addlinespace
\midrule
Observations        &     173,499         &     173,499         &     173,499         &     173,499         \\
Log pseudolikelihood&-53{,}297.314         &-53{,}142.540         &-52{,}476.040         &-53{,}421.087         \\
Pseudo \ensuremath{R\sp{2}}&       0.082         &       0.085         &       0.096         &       0.080         \\
\bottomrule
\end{tabular}
\end{singlespace}
\begin{flushleft}\footnotesize \textit{Notes:} Dependent variable in all columns is \ensuremath{\text{Click}\sb{t+1}}. Each column augments the main MSM specification by interacting \ensuremath{\text{Click}\sb{t}} with a candidate moderator. All models include next-campaign fixed effects and use the same trimmed stabilized IPTW as in Table~\ref{tab:mainMSM}. Omitted categories are faculty (role), the lowest tenure quartile, and zero prior clicks. \end{flushleft}
\end{table}
 
\clearpage

\subsection{Consecutive-campaign restriction}

Table~\ref{tab:ecConsecutive} re-estimates the main MSM outcomes restricting to transitions where the employee's next exposure occurs in the immediately subsequent global campaign (no skipped campaigns).

\begin{table}[htbp]
\centering
\caption{Main MSM estimates, consecutive-campaign transitions}
\label{tab:ecConsecutive}
\def\sym#1{\ifmmode^{#1}\else\(^{#1}\)\fi}
\begin{tabular}{lccc}
\toprule
                    &\multicolumn{1}{c}{(1)}&\multicolumn{1}{c}{(2)}&\multicolumn{1}{c}{(3)}\\
                    &\multicolumn{1}{c}{Click\ensuremath{\sb{t+1}}}&\multicolumn{1}{c}{Report\ensuremath{\sb{t+1}}}&\multicolumn{1}{c}{Safe handling\ensuremath{\sb{t+1}}}\\
\midrule
Click\ensuremath{\sb{t}}&       0.842\sym{***}&       0.262\sym{***}&       0.272\sym{***}\\
                    &     (0.028)         &     (0.039)         &     (0.042)         \\\addlinespace
Constant            &      -1.674\sym{***}&      -4.149\sym{***}&      -4.263\sym{***}\\
                    &     (0.046)         &     (0.125)         &     (0.132)         \\\addlinespace
\midrule
Observations        &     157,156         &     157,156         &     157,156         \\
Log pseudolikelihood&-49{,}143.657         &-33{,}947.012         &-30{,}966.706         \\
Pseudo \ensuremath{R\sp{2}}&       0.079         &       0.093         &       0.098         \\
\bottomrule
\end{tabular}
\begin{flushleft}\footnotesize
\textit{Notes:} Sample restricted to transitions where the employee's next    observed exposure occurs in the immediately subsequent global campaign    (no skipped campaigns). All models include next-campaign fixed effects,    use trimmed stabilized IPTW, and cluster standard errors by employee.    \sym{*} \ensuremath{p<0.10}; \sym{**} \ensuremath{p<0.05};    \sym{***} \ensuremath{p<0.01}.
\end{flushleft}                                                                  
\end{table}
 
\subsection{Similarity robustness and marginal predictions}

Table~\ref{tab:ecSimilarityRobust} reports two robustness checks for similarity-based moderation: (i) adding a control for the log time gap between exposures in the Jaccard interaction model, and (ii) replacing Jaccard with a simple matching coefficient (SMC) that counts both matching presences and matching absences in the cue profile.

\begin{table}[htbp]\centering
\caption{Robustness checks for contextual transfer: gap control and alternative similarity measure.}
\label{tab:ecSimilarityRobust}
\footnotesize
\def\sym#1{\ifmmode^{#1}\else\(^{#1}\)\fi}
\begin{tabular}{l*{4}{c}}
\toprule
                    &\multicolumn{1}{c}{(1)}&\multicolumn{1}{c}{(2)}&\multicolumn{1}{c}{(3)}&\multicolumn{1}{c}{(4)}\\
                    &\multicolumn{1}{c}{Click\ensuremath{\sb{t+1}}}&\multicolumn{1}{c}{Report\ensuremath{\sb{t+1}}}&\multicolumn{1}{c}{Click\ensuremath{\sb{t+1}}}&\multicolumn{1}{c}{Report\ensuremath{\sb{t+1}}}\\
\midrule
Click\ensuremath{\sb{t}}&       0.798\sym{***}&       0.447\sym{***}&       0.752\sym{***}&       0.603\sym{***}\\
                    &     (0.048)         &     (0.083)         &     (0.054)         &     (0.102)         \\\addlinespace
Cue similarity (Jaccard)&       0.811\sym{***}&       1.793\sym{***}&                     &                     \\
                    &     (0.218)         &     (0.635)         &                     &                     \\\addlinespace
Click\ensuremath{\sb{t}} \ensuremath{\times} Cue similarity (Jaccard)&       0.200\sym{**} &      -0.465\sym{***}&                     &                     \\
                    &     (0.081)         &     (0.165)         &                     &                     \\\addlinespace
Cue similarity (SMC)&                     &                     &       0.472\sym{***}&       0.518\sym{*}  \\
                    &                     &                     &     (0.148)         &     (0.305)         \\\addlinespace
Click\ensuremath{\sb{t}} \ensuremath{\times} Cue similarity (SMC)&                     &                     &       0.240\sym{***}&      -0.653\sym{***}\\
                    &                     &                     &     (0.081)         &     (0.167)         \\\addlinespace
Log days to next exposure&      -0.100\sym{***}&      -0.188\sym{***}&                     &                     \\
                    &     (0.028)         &     (0.055)         &                     &                     \\\addlinespace
Constant            &      -2.283\sym{***}&      -4.991\sym{***}&      -2.295\sym{***}&      -4.355\sym{***}\\
                    &     (0.230)         &     (0.580)         &     (0.155)         &     (0.328)         \\\addlinespace
\midrule
Observations        &     173,499         &     173,499         &     173,499         &     173,499         \\
Log pseudolikelihood&-53{,}412.980         &-35{,}264.124         &-53{,}420.913         &-35{,}267.528         \\
Pseudo \ensuremath{R\sp{2}}&       0.080         &       0.101         &       0.080         &       0.101         \\
\bottomrule
\end{tabular}
\begin{flushleft}\footnotesize
\textit{Notes:} Columns (1) and (2) replicate the contextual-transfer models    with an added control for the log number of days between exposures    \ensuremath{t} and \ensuremath{t+1}. Columns (3) and (4) replace Jaccard    similarity with a simple matching coefficient (SMC) that counts both matching    presences and matching absences of cue indicators. All models use trimmed    stabilized IPTW and include next-campaign fixed effects; standard errors are    clustered by employee.    \sym{*} \ensuremath{p<0.10}; \sym{**} \ensuremath{p<0.05};    \sym{***} \ensuremath{p<0.01}.
\end{flushleft}                                                                  
\end{table}

\clearpage

\bibliographystyle{plainnat} 
\bibliography{references_ec}